 \documentclass[pra,12pt,
longbibliography, superscriptaddress, aps,showpacs]{revtex4-2}

\usepackage{graphicx}
\usepackage{amssymb}

\pdfoutput=1
\usepackage[latin1]{inputenc}
\usepackage{amsmath}
\usepackage{amssymb}
\usepackage{color}

\usepackage{mathtools}
\usepackage{float}
\usepackage[T1]{fontenc}

\def\lbCap{\protect\linebreak}

\def\av#1{\langle#1\rangle}

\def\tder#1{\frac{d#1}{dt}}
\def\eps{\varepsilon}
\def\nup{\nu^\prime}

\def\aop{\hat{a}}
\def\adop{\hat{a}^\dagger}
\def\awop{\tilde{a}}
\def\awdop{\tilde{a}^\dagger}
\def\bop{\hat{b}}
\def\bdop{\hat{b}^\dagger}
\def\bwop{\tilde{b}}
\def\bwdop{\tilde{b}^\dagger}

\def\thetap{\theta^\prime}

\def\Hop{\hat{H}}
\def\Bop{\hat{B}}
\def\Bdop{\hat{B}^\dagger}

\def\xiop{\hat{\xi}}

\def\xiwop{\tilde{\xi}}

\def\thetap{{\theta^\prime}}

\def\nn{\nonumber}

\newcommand{\lsz}{\left[}
\newcommand{\rsz}{\right]}
\newcommand{\lk}{\left(}
\newcommand{\rk}{\right)}
\newcommand{\lka}{\left\{}
\newcommand{\rka}{\right\}}

\usepackage{setspace}
\usepackage[font={small,it,bf},labelfont=bf]{caption}
\usepackage{hyperref}
\def\dul#1{\underline{\underline{#1}}}
\def\ul#1{\underline{#1}}

\def\frop#1{\hat{\tilde{#1}}}
\def\av#1{\langle#1\rangle}

\def\refeq#1{{\hyperref[#1]{(\ref*{#1})}}}
\def\reffig#1{{\hyperref[#1]{Fig. \ref*{#1}}}}
\def\refsec#1{{\hyperref[#1]{Sec. \ref*{#1}}}}
\def\refapp#1{{\hyperref[#1]{App. \ref*{#1}}}}
\def\refno#1{{\hyperref[#1]{\ref*{#1}}}}

\begin{document}

\title{Study of the EPR-type entanglement properties of a NDPA with time-delayed coherent feedback}

\author{Nikolett N\'emet}
\email{nemet.nikolett@wigner.hu}
\affiliation{Department of Quantum Optics and Quantum Information, Institute of Solid State Physics,
             Wigner Research Center for Physics, Budapest, Hungary}

\author{Tam\'as Kiss}
\affiliation{Department of Quantum Optics and Quantum Information, Institute of Solid State Physics,
             Wigner Research Center for Physics, Budapest, Hungary}
\author{Scott Parkins}
\affiliation{Department of Physics, 
             University of Auckland, Auckland and\\
             The Dodd-Walls Centre for Photonic and Quantum Technologies, New
             Zealand}
\vspace{-4cm}
\begin{abstract}
An open non-degenerate parametric oscillator (NDPO) is studied below threshold in the undepleted pump regime with time-delayed coherent feedback (TDCF). The figure of merit is the two-mode squeezing spectrum measuring the strength of continuous-valued Einstein-Podolsky-Rosen-type (EPR-type) entanglement between the down-converted modes in the output field 
Exploring different areas of the parameter space reveals the possibility of significantly enhanced entanglement with tunable characteristic frequency. These resonance-like features are closely related to the unique dynamics emerging as a result of delayed feedback. The system is very sensitive to phase matching, which can be adjusted by various parameters of the setup. Further advantage of the current scheme is that significant enhancement of entanglement can be achieved even with finite extraneous losses.
\end{abstract}

\maketitle
\newpage

\section{Introduction}

Quantum theory entails a collection of unique, classically non-intuitive phenomena. Entanglement, for example, infer
subsystems 
showing instantaneous influence on each other even when separated in space. These characteristics are due to the formation of non-local, non-classical correlations, whose strength is independent of the physical distance between the constituents \cite{Horodecki2007}.Nevertheless, these systems are never fully isolated; these correlations might also be influenced by additional degrees of freedom of the environment 
 \cite{Eisert2006}. Applications already exist; entanglement is an essential resource for quantum information processing, quantum computation, and quantum cryptography \cite{NielsenChuangBook2010, Pichler2017}, and it also improves the sensitivity in precision measurement setups \cite{QsensingRev2017}.

Macroscopic detection of quantum mechanical features relies on the preservation of quantum coherence within the system \cite{Glauber1963}. One of the biggest hindrances to the up-scaling of today's quantum technologies is environment-induced decoherence. A whole research field is dedicated to the study of open quantum systems \cite{Breuer2002TheTO}, where one of the main objectives is to build the most accurate model of the measurement setup and more specifically of the interaction between the surrounding reservoir and the observed quantum system. These models form the underlying foundations of how decoherence works in a given setup and how its influence is identified or even mitigated in the detected signal.
An approach is to consider the surroundings as a reservoir and its influence on the system as random fluctuations, similar to Brownian motion \cite{Carmichael2003StatisticalMI}. Although investigations mostly focus on the adverse effects of system-reservoir interactions due to input fluctuations, field leaking out of the system, on the other hand, is the source of the purest (quantum) information about the system's present state. This field can be used to study or even control the properties of a setup \cite{Koswara2021,DongPetersenBook2023}.

In control theory a well-known technique for improving the signal-to-noise ratio is to alter the input signal based on the information obtained from the measurement output \cite{Zhang2017}. This methodology has also been applied to quantum systems, where measurement-based feedback is used to maintain system population and/or coherence \cite{Wiseman2010}. Therefore, in this case quantum coherence is not preserved in the feedback loop.

Recent experimental advances allow longer coherence times in waveguides. Thus, implementing {\em coherent} or {\em measurement-free feedback} is becoming more and more feasible \cite{Eschner2001,Wilson2003,Dubin2007,Mabuchi2008}. All-optical, short-delay feedback was applied in an experiment to improve the quantum signal-to-noise ratio \cite{Iida2012}. The results confirmed that circumventing measurements can give a much clearer picture of the internal state of the system as the number of interacting systems is minimized, leading to more efficient control \cite{Jacobs2014}. Nevertheless, most theoretical works have focused on short coherent feedback loops, where the propagation time delay is negligible, as in that case the setup is less prone to decoherence and loss processes \cite{Dong2010,DongPetersenBook2023}.

Research fields such as precision measurement or the study of quantum networks bring in two-point interactions between fields carrying information about the system state at different time instants. These delayed interactions enable substantial alteration of the dynamical properties of the system \cite{Geffert2015,Jaurigue2017}. In this case time-reversal symmetry is broken by the unidirectional propagation, resulting in a feedback loop instead of an additional resonator. Assume, for example, that a field including time-delayed coherent feedback (TDCF) of the system output field interferes destructively with the input field. 
Then one obtains the open quantum system analogue of Pyragas-type feedback \cite{Kabuss2016}. This type of feedback was originally introduced as a method to stabilize unstable periodic orbits in classical chaotic dynamics by tuning the feedback-loop length to resonate with the target system's oscillations \cite{Pyragas1992}. It is frequently applied in contemporary studies \cite{Geffert2015, Jaurigue2017}, especially in laser theory. Stabilization of, or faster convergence to certain steady states can also, in principle, be obtained in this way for a single two-level system in front of a mirror \cite{Droenner2019}, for optomechanical setups \cite{Naumann2014}, or even for the Dicke model \cite{Grimsmo2014}. 

The above-described studies paved the way for the application of feedback loops as ``controllers'' of the original ``plant'' quantum system. Adding a controller to a setup mathematically translates into a dynamical-landscape extension, with the dimensions corresponding to quantities such as the ratio of the returning and initial input fields, accumulated phase shift, or time delay associated with the loop length. Carefully chosen sets of these control parameters may induce substantially different attractors from those of the original system. Persistent oscillations, for example, may emerge as a result of the introduction of delayed interactions \cite{Carmele2013,Kabuss2015,Kraft2016,Nemet2016}.

The present study also shows that more than one parameter can act as a control parameter, the tuning of which may shift the dynamical behavior from the original towards the new regime. Combining the obtained dynamical variety with a quantum mechanical description, enhanced quantum squeezing \cite{Mabuchi2008,Gough2009, Iida2012, Crisafulli2013, Kraft2016,Nemet2016}, stronger entanglement \cite{Hein2014,Hein2015,Hein2016a}, or even conserved coherence in a thermal environment \cite{Nemet2019} have been found. Very recently, an experiment has been described using TDCF control of a vibrating membrane inside a one-sided resonator to drive it towards a target state \cite{Schmid2022,Ernzer2023}.

\begin{figure}[t!]
	\centering
	\vglue -.3 cm
	\includegraphics[width=.9\textwidth]{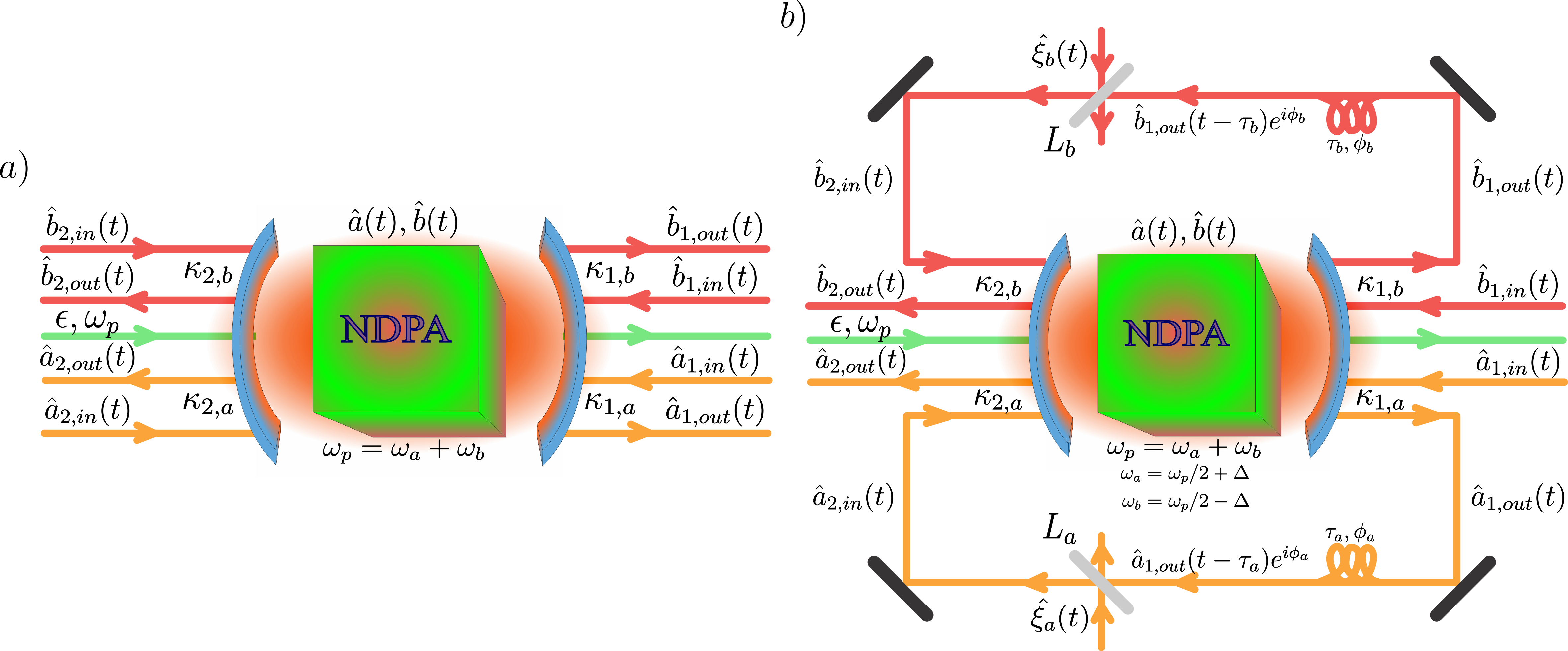}
	\vglue -.2 cm
	\caption{\linespread{1}{Schematic setup. A non-linear crystal performing parametric down-conversion (with non-linearity $\chi$, mode frequencies $\omega_{a,b}$, with $\omega_a-\omega_b=2\Delta$) on the incoming pump photons (with amplitude $\epsilon$ and frequency $\omega_p$) is placed in an initially empty cavity (with decay rates $\kappa_{\{1,2\},\{a,b\}}$). Pump depletion is neglected, the mode-coupling pump parameter is defined as $\varepsilon=\epsilon\chi$. a) Regular NDPO without feedback. 
			b) NDPO with TDCF: The output fields on the right-hand side are coupled directly back into the input channels of the other side, without performing any measurements. The longer lengths of the feedback loops enable finite time delays ($\tau_{a,b}$) and phase shifts ($\phi_{a,b}$). Losses in the feedback loops are also modeled via beam splitters ($L_{a,b}$), allowing for vacuum noise ($\xiop_{1,2}$) in the feedback loop. }}\label{fig:setup}
	\vglue -.5 cm
\end{figure}

This paper focuses on the effects of TDCF as a controller for the non-degenerate parametric oscillator (NDPO); more specifically, on the influence it has on the output continuous variable entanglement properties \cite{Collett1987,Drummond1990}. 
The simplest realization of an NDPO is a crystal with $\chi^{(2)}$ non-linearity performing parametric down-conversion inside an optical cavity, converting the pump mode into two distinct field modes that are supported by the resonator (see \reffig{fig:setup}.$a)$). Output quadrature detection of the two down-converted modes shows strong correlations or anti-correlations depending on the quadrature phase angle \cite{Ou1992}. Although entanglement of continuous variables, such as quadratures of light, is already readily available by non-linear optical processes, its quality is generally limited compared to its discrete-variable counterparts \cite{Braunstein2003,Braunstein2005}.

In order to enhance the performance of entanglement generation, generally larger driving power is required as the best performance can be obtained at the threshold for parametric oscillation \cite{Collett1987,Drummond1990}. Increasing the driving strength above a certain level, however, triggers adverse processes, such as heating, scattering, and other uncontrolled non-linear effects. In order to overcome these intrinsic limitations to the system's performance, cascaded \cite{He2007,Shi2016a} and feedback setups \cite{Li2006,Ke2007,Yan2011,Zhou2015}) have been proposed. Enhanced entanglement has been shown for feedback setups over a broader range of frequencies \cite{He2007,Shi2016a,Yan2011,Zhou2015} and the potential for multipartite entanglement generation has also been proposed. Nevertheless, the corresponding experiment \cite{Zhou2015} considers only two modes, as for more modes, non-trivial phase matching of multiple non-linearities is required \cite{Pfister2004}.

Closely-related studies have focused on the properties of a special case of this NDPO, the degenerate parametric amplifier (DPA), where the two down-converted field modes coincide. In that case, enhanced squeezing was shown at shifted frequencies and reduced threshold pump powers \cite{Kraft2016,Nemet2016}. This is due to the previously mentioned versatile dynamics that translates into a unique stability landscape, a characteristic feature of delayed feedback. The analytical description enables simple tuning of selected parameters in order to explore the obtained multidimensional phase space.

As the NDPA is a relatively straightforward generalization of the DPA, similar features are to be expected for the two-mode squeezing spectrum, which is known to be a direct measure of continuous variable entanglement \cite{Caves1985,Schumaker1985}. 
Potential experimental platforms include circuit quantum electrodynamic (circuit-QED) systems \cite{You2011}, where, for example, electromagnetically induced transparency can be used to ``slow down'' light, enabling longer, tunable delays over a shorter distance. Intrinsic diffusion in these waveguides, on the other hand, results in larger loss rates during signal propagation. Topological systems are also promising due to symmetry-protected paths of signal propagation \cite{TopInsHun2015,Parto2020}. An optical resonator side-coupled with an optical fiber in a two-point contact scheme is another tunable-delay option. Using nanofiber cavities, the Fiber Bragg grating mirrors provide tunability of the resonator outcoupling rate \cite{Kato2015}.

This paper is structured as follows; Section II introduces the underlying mathematical model with physically reasonable assumptions. Section III showcases the entanglement characteristics of the output field in a setup with and without feedback. It also connects the findings with the parameter-dependence of the stability landscape. Section IV shows the properties that stem from the different parameter sets for the two down-converted modes distinguishing this system from the DPA. The final section concludes the paper by summarizing its findings.

\section{Quantum Mechanical Model}

The setup considered in this Paper is a TDCF scheme for an NDPA embedded in an optical resonator, also depicted in \reffig{fig:setup}. The obtained NDPO is considered to operate below the oscillation threshold, thus, it is adequate to treat the pump semiclassically, i.e., as a continuous coherent source with amplitude $\epsilon$ that is constant (i.e., undepleted). The down-converted modes $\hat a$ and $\hat b$ are distinguishable either by their polarizations or their resonance frequencies. In the latter case, the difference between the two mode frequencies is expressed as $\omega_a-\omega_b = 2\Delta$. The Hamiltonian in a frame rotating at half the pump frequency ($\omega_p/2$) has the form \cite{Whalen2015}
\begin{align}
	\label{eq:HS}
	\Hop_S &= \Delta\lk\adop\aop-\bdop\bop\rk + i\lk \eps\adop\bdop-\eps^*\aop\bop\rk,
\end{align}
where $\eps=\chi\epsilon\equiv |\eps|e^{i\theta}$ is referred to as the pump parameter throughout the paper and it incorporates both the material's intrinsic non-linearity ($\chi$) and the coherent pump amplitude ($\epsilon$) (also shown in \reffig{fig:setup}). 

Two separate loops around the system create a link between the output and input fields of the two distinct down-converted modes, establishing a double-feedback scheme. The output fields via one of the mirrors of the cavity are directly coupled back into the input channels of the other mirror, without implementing any measurements. This requires unidirectional propagation with long coherence times within the feedback channel or waveguide.

The lengths of the feedback loops translate into finite time delays $\tau_a, \tau_b$ that are, in the most general case, different from each other. The mirror transmission rates for the cavity on each side (1 and 2) for each down-converted mode ($\alpha\in\lka a,b\rka$) are expressed within the finite decay rates of the corresponding modes, $\kappa_{j,\alpha}$ ($j\in\{1,2\},\alpha\in\{a,b\}$). 
The phase shifts - emerging due to, for example, propagation in the feedback loops, phase shifters, reflections and scattering off the elements of the setup - are collected in respective phase factors $\phi_{\alpha}$ ($\alpha\in\{a,b\}$). Losses as well as additional vacuum noise contributions in each loop are modeled by a beam-splitter interaction. 
Throughout this paper we consider only vacuum inputs for $\aop_{1,in}(t)$ and $\bop_{1,in}(t)$, as well as for the loss channels $\xiop_\alpha(t)$ ($\alpha\in\{a,b\}$). 
Note that unless the polarizations of the two modes are orthogonal, the frequency difference for non-overlapping down-converted modes has to satisfy 
\begin{align}
	\label{eq:Delta_cond}
	2\Delta&\gg\kappa_a,\kappa_b,
\end{align}
for the distinguishability of these modes, where $\kappa_\alpha=\kappa_{1,\alpha}+\kappa_{2,\alpha}$ (with $\alpha\in\lka a,b\rka$) are the respective decay rates for the corresponding cavity modes.
\subsection{Time evolution}
In our model TDCF is implemented with the output fields on the right-hand side of the cavity coupling back after a certain time into the input channels of the left-hand side, for each mode. This delayed re-entering of quantum information about the system's past state induces non-Markovian evolution for the system, which can be described by the following quantum Langevin equations (\refeq{eq:gen_Heis} from \refapp{app:Tev_der}), 
\begin{align}
	\label{eq:aop_tder}
	\tder{\aop(t)} = -\lk \kappa_a + i\Delta\rk\aop(t) + \eps\bdop(t)-\sqrt{2\kappa_{1,a}}\aop_{1,in}(t)-\sqrt{2\kappa_{2,a}}\aop_{2,in}(t),\\
	\label{eq:bop_tder}
	\tder{\bop(t)} = -\lk \kappa_b - i\Delta\rk\bop(t) + \eps\adop(t) -\sqrt{2\kappa_{1,b}}\bop_{1,in}(t)-\sqrt{2\kappa_{2,b}}\bop_{2,in}(t).
\end{align}
The non-Markovian character comes from the time-delayed output field contributions at the input channels $\hat{a}_{2,in}$ and $\hat{b}_{2,in}$, 
\begin{align}
	\label{eq:a2in}
	\aop_{2,in}(t) &= \sqrt{1-L_a}e^{i\phi_a}\aop_{1,out}(t-\tau_a)+\sqrt{L_a}\xiop_a(t)\nn\\
	&=\sqrt{1-L_a}e^{i\phi_a}\lsz\aop_{1,in}(t-\tau_a)+\sqrt{2\kappa_{1,a}}\aop(t-\tau_a)\rsz+\sqrt{L_a}\xiop_a(t),\\   
	\label{eq:b2in}
	\bop_{2,in}(t) &= \sqrt{1-L_b}e^{i\phi_b}\bop_{1,out}(t-\tau_b)+\sqrt{L_b}\xiop_b(t)\nn\\
	&=\sqrt{1-L_b}e^{i\phi_b}\lsz\bop_{1,in}(t-\tau_b)+\sqrt{2\kappa_{1,b}}\bop(t-\tau_b)\rsz+\sqrt{L_b}\xiop_b(t),
\end{align}
where we have used the input-output theorem, and imperfect propagation in the feedback loop is modeled via a beam splitter, describing both noise and loss processes. The noise contribution is modeled as vacuum fluctuation input ($\xiop_\alpha$) mixing with the original feedback field. A more explicit form of the equations is obtained by collecting all the vacuum fluctuation input terms as a single noise term,
\begin{align}
	\label{eq:aop_der}
	\tder{\aop(t)} = -\lk \kappa_a + i\Delta\rk\aop(t) + \eps\bdop(t)-e^{i\phi_a}k_a\aop(t-\tau_a)-\sqrt{2\kappa_a}\aop_{in}(t),\\
	\label{eq:bop_der}
	\tder{\bop(t)} = -\lk \kappa_b - i\Delta\rk\bop(t) + \eps\adop(t) -e^{i\phi_b}k_b\bop(t-\tau_b)-\sqrt{2\kappa_b}\bop_{in}(t),
\end{align}
where $k_\alpha =2\sqrt{\kappa_{1,\alpha}\kappa_{2,\alpha}(1-L_\alpha )}$ ($\alpha =a,b$) is the feedback strength for the system observables, which depends on the loss rate $L_\alpha$ in the feedback channel. A competition is present between the corresponding vacuum field and feedback terms, where in the extreme case of $100\%$ loss, the original NDPO set-up is recovered, as in that case the feedback field is entirely replaced by vacuum noise. This can also be identified from \refeq{eq:a2in} and\refeq{eq:b2in}. The noise term incorporates the reservoir part of the delayed field, introducing the time delay as a characteristic timescale (color) of the reservoir:
\begin{align}
	\label{eq:ain}
	\aop_{in}(t) &= \frac{1}{\sqrt{2\kappa_a}}\lsz\sqrt{2\kappa_{1,a}}\aop_{1,in}(t)+\sqrt{2\kappa_{2,a}}\aop'_{2,in}(t)\rsz,\\
	\label{eq:ap2in}
	\aop'_{2,in}(t)&=\sqrt{1-L_a}e^{i\phi_a}\aop_{1,in}(t-\tau_a)+\sqrt{L_a}\xiop_a(t),\\
	\label{eq:bin}
	\bop_{in}(t) &= \frac{1}{\sqrt{2\kappa_b}}\lsz\sqrt{2\kappa_{1,b}}\bop_{1,in}(t)+\sqrt{2\kappa_{2,b}}\bop'_{2,in}(t)\rsz,\\
	\label{eq:bp2in}
	\bop'_{2,in}(t)&=\sqrt{1-L_b}e^{i\phi_b}\bop_{1,in}(t-\tau_b)+\sqrt{L_b}\xiop_b(t).
\end{align}
Thus, white noise is replaced by colored noise, the environment obtains a memory character.
\subsection{Output field spectrum}

\noindent In the undepleted pump regime, below the threshold of parametric oscillation, the dynamics of the NDPO is described by a set of linear (operator) equations of motion. The same is true when TDCF is applied, despite the time-nonlocal character that may induce new types of solutions. In frequency space the time delay only serves as an extra, frequency-dependent phase factor, which allows for a simple, closed-form description of entanglement properties. Note that the adjoint operator in that case moves to the other side of the spectrum, such that
\begin{align}
	\hat{\tilde{O}}(\nu)&=\frac{1}{\sqrt{2\pi}}\int_{-\infty}^\infty e^{i\nu t}\hat{O}(t)dt, &	\lk\frop{O}(\nu)\rk^\dagger&=\frop{O}^\dagger(-\nu).
\end{align}
Applying this to equations \refeq{eq:aop_der} and \refeq{eq:bop_der} results in
\begin{align}
	\awop(\nu) &= \frac{1}{\Lambda_{ba}(\nu)}\lsz d_{+b}(\nu)\sqrt{2\kappa_a}\awop_{in}(\nu)+\eps\sqrt{2\kappa_b}\bwdop_{in}(-\nu)\rsz,\\
	\bwop(\nu) &= \frac{1}{\Lambda_{ab}(\nu)}\lsz d_{+a}(\nu)\sqrt{2\kappa_b}\bwop_{in}(\nu)+\eps\sqrt{2\kappa_a}\awdop_{in}(-\nu)\rsz \text{, with}
\end{align}
\vspace{-1.5cm}
\begin{align}
	d_{\pm,a}(\nu ) &= \kappa_a-i\lk\nu\pm\Delta\rk+k_ae^{i(\nu\tau_a\mp\phi_a)},&
	d_{\pm,b}(\nu ) &= \kappa_b-i\lk\nu\mp\Delta\rk+k_be^{i(\nu\tau_b\mp\phi_b)},
\end{align}
where the feedback contribution is incorporated in the input fields of \refeq{eq:ain} and \refeq{eq:bin} together with loss-induced vacuum noise,
\begin{align}
	\awop_{in} (\nu) &= \frac{1}{\sqrt{2\kappa_a}}\lsz f_{1,a}(\nu)\awop_{1,in}(\nu)+\sqrt{2\kappa_{2,a}L_a}\,\xiwop_{a}(\nu)\rsz,
	\\
	\bwop_{in} (\nu) &= \frac{1}{\sqrt{2\kappa_b}}\lsz f_{1,b}(\nu)\bwop_{1,in}(\nu)+\sqrt{2\kappa_{2,b}L_b}\,\xiwop_{b}(\nu)\rsz,
\end{align}
where the Markovian and non-Markovian contributions of the original vacuum noise input fields are combined as 
\begin{align}
	f_{j,\alpha}(\nu) &= \frac{1}{\sqrt{2\kappa_{j,\alpha}}}\lsz 2\kappa_{j,\alpha} + k_\alpha e^{i(\nu\tau_\alpha+\phi_\alpha)}\rsz , & j&\in\{1,2\}, \alpha\in\{a,b\}.
\end{align}
The monitored output field is obtained using the above expressions in the input-output theory,
\begin{align}
	\label{eq:aout}
	\frop{a}_{2,out}(\nu) =  \frop{a}_{2,in}(\nu)+\sqrt{2\kappa_{2,a}}\frop{a}=\frac{1}{\Lambda_{ba}(\nu)}&\lka D_a(\nu)\frop{a}_{1,in}(\nu)+E_a(\nu)\frop{\xi}_a(\nu)+\right. \\
	&\left.\eps f_{2,a}(\nu)\lsz f_{1,b}^*(-\nu)\frop{b}^\dagger_{1,in}(-\nu)+\sqrt{2\kappa_{2,b}L_b}\frop{\xi}^\dagger_b(-\nu)\rsz\rka,\nn\\
	\label{eq:bout}
	\frop{b}_{2,out}(\nu)= \frop{b}_{2,in}(\nu)+\sqrt{2\kappa_{2,b}}\frop{b}=\frac{1}{\Lambda_{ab}(\nu)}&\lka D_b(\nu)\frop{b}_{1,in}(\nu)+E_b(\nu)\frop{\xi}_b(\nu)+\right. \\
	&\left.\eps f_{2,b}(\nu)\lsz f_{1,a}^*(-\nu)\frop{a}^\dagger_{1,in}(-\nu)+\sqrt{2\kappa_{2,a}L_a}\frop{\xi}^\dagger_a(-\nu)\rsz\rka,\nn
\end{align}
with the expressions
\begin{align}
	\label{eq:Lamab}
	\Lambda_{ab}(\nu) &= |\eps|^2-d_{+a}(\nu)d_{-b}(\nu),\\
	\Lambda_{ba}(\nu) &= |\eps|^2-d_{+b}(\nu)d_{-a}(\nu),
	\end{align}\begin{align}
	\label{eq:D_a}
	D_a(\nu) &= \sqrt{1-L_a}e^{i(\nu\tau_a+\phi_a)}\lka|\eps|^2+\lsz\kappa_a+i(\nu-\Delta)\rsz d_{+b}(\nu)\rka+2\sqrt{\kappa_{1,a}\kappa_{2,a}}d_{+b}(\nu),\\
	\label{eq:D_b}
	D_b(\nu) &= \sqrt{1-L_b}e^{i(\nu\tau_b+\phi_b)}\lka|\eps|^2+
	\lsz\kappa_b+i(\nu+\Delta)\rsz d_{+a}(\nu)\rka+2\sqrt{\kappa_{1,b}\kappa_{2,b}}d_{+a}(\nu),\\
	\label{eq:E_a}
	E_a(\nu) &= \sqrt{L_a}\lka |\eps|^2 + 
	\lsz\kappa_{2,a}-\kappa_{1,a}+i(\nu-\Delta)\rsz d_{+b}(\nu)\rka,\\
	\label{eq:E_b}
	E_b(\nu) &= \sqrt{L_b}\lka |\eps|^2 + 
	\lsz\kappa_{2,b}-\kappa_{1,b}+i(\nu+\Delta)\rsz d_{+a}(\nu)\rka.
\end{align}
Notice that, contrary to the DPA case, the presence of the non-linearity introduces cross-correlations between the different modes. Moreover, the induced correlations are present in the time domain as well, as the present and delayed fields of one mode couple to the present and delayed fields of the other mode.
Thus, even when feedback is considered just for one mode in terms \refeq{eq:D_a} and \refeq{eq:D_b}, the other still senses the introduced memory effects due to the intrinsic signal-idler entanglement in the system.

The coefficients \refeq{eq:E_a} and \refeq{eq:E_b} represent the noise terms that only play a role when imperfections are included via a finite loss rate of the feedback loop ($L_{a,b}$). Interestingly, when the two modes are resonant, for a cavity that has the same decay rate for both directions, i.e., is symmetric, the noise contribution grows linearly with the pump field intensity and the time delay does not play any role. 
\section{EPR-type Entanglement}

\noindent The NDPO can be described by a simple model using the undepleted pump approximation \refeq{eq:HS}. The emerging linear dynamics allows for an analytical description of the continuous variable EPR-type entanglement between the two down-converted modes (\refapp{app:2ModeSqueez}) \cite{Collett1987,Drummond1990}. A common measure to use 
is the noise suppression in a generalized quadrature combining the effect of the two
modes. Thus, this two-mode squeezing spectrum 
\cite{Caves1985,Schumaker1985} is monitored as an entanglement witness based on a heterodyne (in the case of non-overlapping modes in frequency) or coupled homodyne detection scheme (for frequency-degenerate modes with orthogonal polarizations) \cite{Drummond1990}.

In a one-sided cavity without feedback, considering orthogonally polarized down-converted modes of the same frequency, the noise spectrum of the output modes in \refeq{eq:chi_nu} has the following form \cite{Collett1987},
\begin{align}
	\chi_{\thetap,out}^\text{no fb}(\nu)&=\lka 1+4|\eps|\kappa\lsz\frac{\cos^2{\lk\frac{\thetap}{2}\rk}}{\lk\kappa-|\eps|\rk^2+\nu^2}-\frac{\sin^2{\lk\frac{\thetap}{2}\rk}}{\lk\kappa+|\eps|\rk^2+\nu^2}\rsz\rka.
\end{align}
Note that, compared to \cite{Collett1987}, we have defined $\eps$ with an extra minus sign. 
For the phase quadrature described by $\thetap=\pi$ this expression simplifies to 
\begin{align}
	\label{eq:nofb_eqfreq_ref}
	\chi_{\pi,out}^\text{no fb}(\nu)=1-\frac{4|\eps|\kappa}{\lk\kappa+|\eps|\rk^2+\nu^2},
\end{align}
which vanishes on resonance at the oscillation threshold ($|\eps|=\kappa$), corresponding to, in principle, perfect entanglement between the quadratures of the two down-converted modes, in the undepleted pump approximation. 
Note that this simple picture is possible due to, among other factors, the lack of pump depletion in the model, and the derived expression gives the same noise power spectrum as the single-mode expression for a DPA.

In contrast, non-identical, well-separated frequencies of the down-converted modes (with identical polarizations) entail a frequency-split noise power spectrum \cite{Collett1987},
\begin{align}
	\label{eq:nofb_spec}
	\chi_{\pi,out}^\text{no fb}(\nu)=1-\frac{4|\eps|\kappa}{\lk\kappa+|\eps|\rk^2+(\nu+\Delta)^2}-\frac{4|\eps|\kappa}{\lk\kappa+|\eps|\rk^2+(\nu-\Delta)^2}.
\end{align}
The two negative terms in (\ref{eq:nofb_spec})
seemingly double the squeezing effect (cf. \refeq{eq:nofb_eqfreq_ref}), but, as the two modes are non-overlapping ($\Delta\gg\kappa$), the contributions of each of the terms are centered around different frequencies ($\nu=\pm\Delta$). Note that in a typical measurement setup, for frequency-degenerate modes with orthogonal polarizations, two local oscillators are used for detection. Meanwhile, in the case of large frequency difference between the down-converted modes compared to the resonator linewidth (but identical polarizations), a single local oscillator suffices, as only one of the terms is relevant for each side of the spectrum \cite{Drummond1990}. 

\subsection{With Feedback}

\noindent The introduction of TDCF alters the entanglement properties of the system, as it introduces a structure, an additional timescale into the setup. The dynamics on this timescale interferes with the original dynamics, and this interference and competition results in a unique two-mode squeezing spectrum. Based on the standard methods in \refapp{app:2ModeSqueez} using equations \refeq{eq:aout} and \refeq{eq:bout}, we can derive the following spectrum:
\begin{align}\label{eq:chi_out}
	\chi_{\thetap,out}(\nu) &= 
	1+\frac{|\eps|}{2}\lka\frac{2\Re\lk M_{ba}(\nu)\rk+N_{ba}(\nu)}{\left|\Lambda_{ab}(\nu)\right|^2}+\frac{2\Re\lk M_{ab}(\nu)\rk+N_{ab}(\nu)}{\left|\Lambda_{ba}(\nu)\right|^2}\rka,
\end{align}
where,
\begin{align}
	M_{ab}(\nu) &= e^{i\thetap}f_{2,b}(-\nu)\lsz D_a(\nu)f_{1,a}^*(\nu)+\sqrt{2L_a\kappa_{2,a}}E_a(\nu)\rsz,
	\\
	M_{ba}(\nu) &= e^{i\thetap}f_{2,a}(-\nu)\lsz D_b(\nu)f_{1,b}^*(\nu)+\sqrt{2L_b\kappa_{2,b}}E_b(\nu)\rsz,
	\\
	N_{ab}(\nu) &= |\eps|\lka\left|f_{2,b}(-\nu)\right|^2\lsz\left|f_{1,a}(\nu)\right|^2+2L_a\kappa_{2,a}\rsz+\left|f_{2,a}(\nu)\right|^2\lsz\left|f_{1,b}(-\nu)\right|^2+2L_b\kappa_{2,b}\rsz\rka,
	\\
	N_{ba}(\nu) &= |\eps|\lka\left|f_{2,a}(-\nu)\right|^2\lsz\left|f_{1,b}(\nu)\right|^2+2L_b\kappa_{2,b}\rsz+\left|f_{2,b}(\nu)\right|^2\lsz\left|f_{1,a}(-\nu)\right|^2+2L_a\kappa_{2,a}\rsz\rka,
\end{align}
and $\thetap=\theta-(\thetap_a+\thetap_b)/2$. Notice that expression (\ref{eq:chi_out}) has the same structure as \refeq{eq:nofb_spec}; the last two terms play the same role in both cases. They correspond to opposite halves of the spectrum, where the contribution proportional to $1/\Lambda_{ab}(\nu)$ is centered around $-\Delta$ [see \refeq{eq:Lamab}].

\begin{figure}[h!]
	\centering
	\includegraphics[width=.9\textwidth]{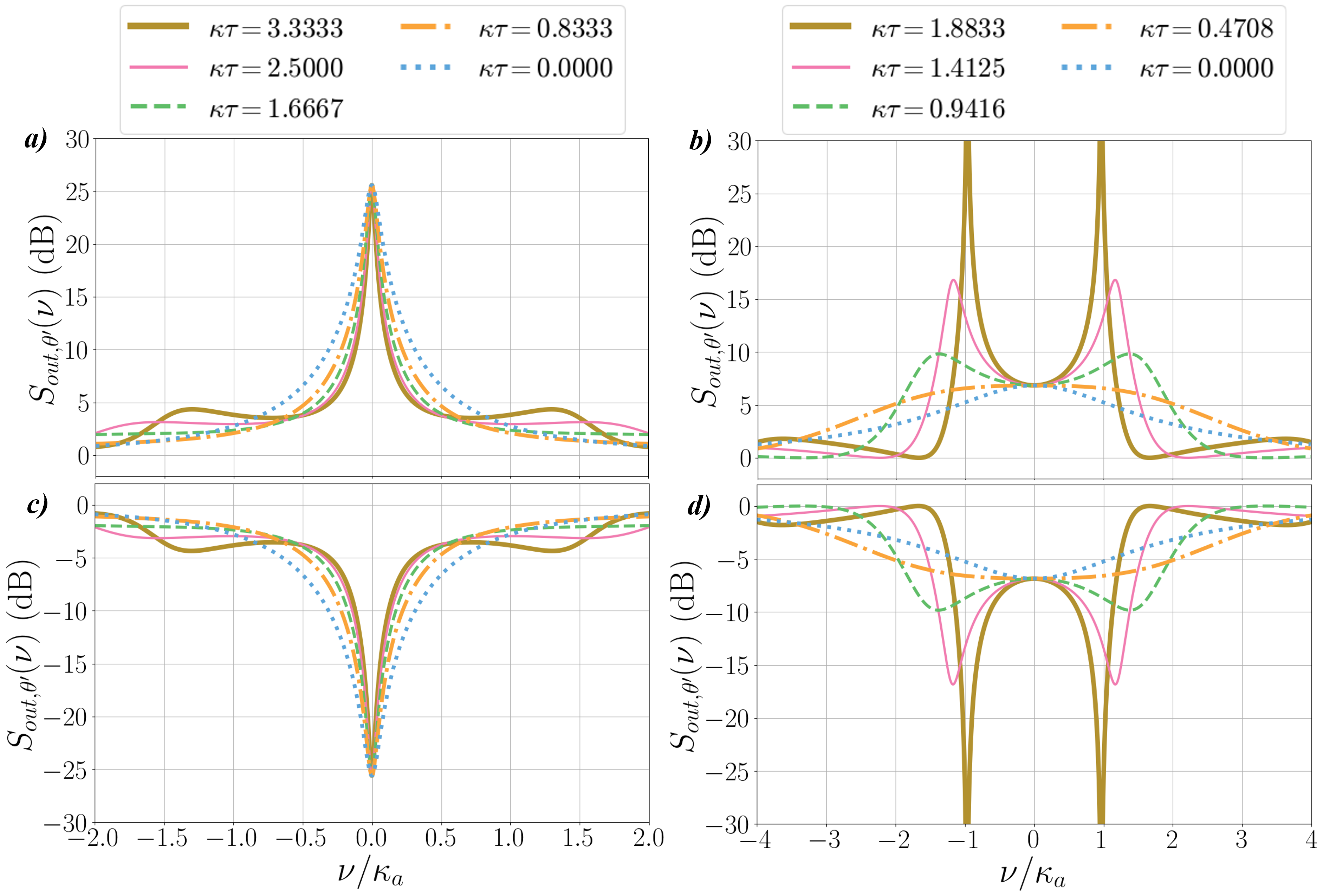}
	\vglue -0.3 cm
	\caption{\linespread{1}Noise power spectra in a) an almost one-sided and b) a symmetric cavity with coherent time-delayed lossless feedback. \lbCap
		Parameters: $\kappa_\alpha=10\cdot2\pi$ MHz, where $\alpha\in\lka a,b\rka,\Delta = 0$ MHz \lbCap
		left: $\phi_\alpha=\pi, |\eps|=0.45\kappa_a, \kappa_{1,\alpha} = 0.933\kappa_\alpha
		$,
		right: $\phi_\alpha=0,|\eps|=0.75\kappa_a, \kappa_{1,\alpha} = \kappa_\alpha/2. 
		$}\label{fig:delays}
	\vglue -.3 cm
\end{figure}

{\bf Pyragas-type feedback}  
is a straightforward theoretical concept, as it matches the phase and the characteristic frequency of the intrinsic periodic solution. Naturally, it was mostly introduced to stabilize unstable periodic orbits. Here, by fine-tuning the time delay corresponding to the feedback loop, the period of the desired solution is matched and, thus, the feedback term cancels. In our case, setting the feedback phase to a specific value $\phi_a=\phi_b=\pi$ and significantly reducing one of the mirror transmissions to obtain an almost one-sided cavity, $\kappa_{1,\alpha}\simeq\kappa_\alpha \lka\alpha \in \lka a,b\rka\rk$, results in a similar dynamics, where past and present fields have the same strengths but opposite signs. 
In this way, they cancel each other at certain values of the delay or when a steady state is reached \cite{Pyragas1992}. The characteristic influence of this type of feedback on the NDPA is similar to that previously found for degenerate modes \cite{Nemet2016}. It results in the following equations of motion,
\begin{align}
	\tder{\aop(t)} = -i\Delta\aop(t) -(\kappa_a-k_a)\aop(t)+\eps\bdop(t)-k_a\lsz\aop(t)-\aop(t-\tau_a)\rsz-\sqrt{2\kappa_a}\aop_{in}(t),\\
	\tder{\bop(t)} = i\Delta\bop(t) -(\kappa_b-k_b)\bop(t)+\eps\adop(t)-k_b\lsz\bop(t)-\bop(t-\tau_b)\rsz-\sqrt{2\kappa_b}\bop_{in}(t).
\end{align}
Notice the reduced effective damping terms due to destructive interference between past and present fields. \reffig{fig:delays}.$a)$ and $c)$ show the obtained anti-squeezing and squeezing spectra, respectively. For longer time delays, enhanced side-peaks (dips) emerge, however the best squeezing always occurs on resonance.


\begin{figure}[h!]
	\centering
	\vglue -2 cm
	\includegraphics[width=.84\textwidth]{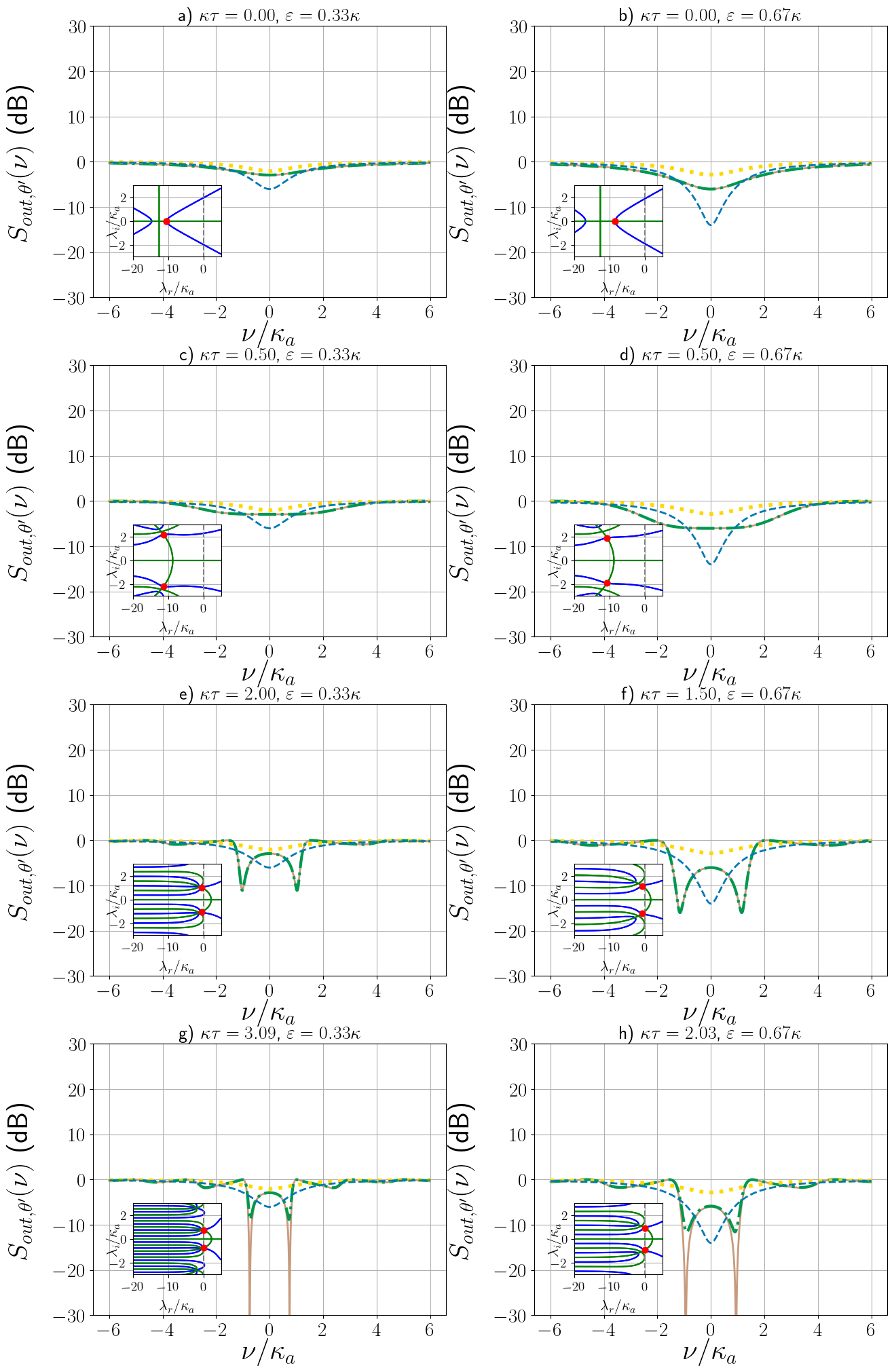}
	\vglue -0.4 cm
	\caption{{\linespread{1}Noise power spectra in a symmetric cavity ($\kappa_{1,\alpha}=\kappa_{2,\alpha}, \alpha\in\{a,b\}$) with lossless (brown solid) and lossy (green dash-dotted) TDCF is compared to the case without feedback in a symmetric ($\kappa_{1,\alpha}=\kappa_{2,\alpha}$, yellow dotted) and one-sided cavity ($\kappa_{1,\alpha}=\kappa_{\alpha}$, blue dashed). The inset shows the solution curves for the stability equations \refeq{eq:stab_a} and \refeq{eq:stab_b} red dots signaling the eigenvalues with the largest real parts.\lbCap Parameters: $\thetap=\pi$, $L_\alpha=5\%$ (for the lossy TDCF case), $\phi_\alpha=0$, $\kappa_\alpha=\kappa$, $\tau_\alpha=\tau$, $|\eps|=\kappa/3$ (left), $|\eps|=2\kappa/3$ (right).}}\label{fig:sq_specs}
	\vglue -.3 cm
\end{figure}

{\bf Phase-matched feedback} --  
where the phases of the output fields $\aop_{1,out}$ and $\bop_{1,out}$ are matched with those of the input fields $\aop_{2,in}$ and $\bop_{2,in}$ (i.e., $\phi_a=\phi_b=0$) -- shows much more variation in its influence on the system. The noise characteristics in this case depend on the relationship between the timescales of cavity damping and feedback. 
In the following we show how the two-mode squeezing spectrum changes as we increase the time delay from 0 to a value that is comparable to, or even longer, than the timescale of the resonator decay.

For short delays with $\kappa_\alpha\tau_\alpha\ll1 \lk\alpha\in\lka a,b\rka\rk$, stabilization of the original solution occurs, and the oscillation threshold is shifted towards larger pump strengths. As the length of the feedback loop increases, side peaks associated with the structure of the feedback reservoir start to emerge gradually. For short delays a broadening of the central peak signals their presence (\reffig{fig:sq_specs}.$b)\&f)$), -- that has also been observed in a cascaded NDPA setup \cite{He2007,Shi2016a}, while for delays longer than the characteristic timescales of the resonator decays, distinct, well-defined side peaks appear. The characteristic frequencies of these peaks tend to a particular resonance frequency, defined by a certain combination of pump strength, mirror transmission rate and time delay. In principle, pure Bell-type entanglement of the two down-converted modes is achievable as a limiting case of the undepleted pump approximation \cite{Nemet2016}. The exact expressions for the corresponding frequencies and time delays can be written as
\begin{align}
	\label{eq:nu_c}
	\nu_c &= \pm\sqrt{k_a^2-(\kappa_a-|\eps|)^2},
	&\tau_{a,c} &= \frac{\arccos{\lk\frac{|\eps|-\kappa_a}{k_a}\rk}}{\nu_c}.
\end{align}
These latter solutions are unique to time-delayed systems, as for an NDPA in which the down-converted modes are resonant, the best two-mode squeezing always occurs on resonance.

The characteristic tendencies of the two-mode anti-squeezing and squeezing spectra as a function of the feedback delay can be seen in \reffig{fig:delays} $b)$ and $d)$. These figures also show that, contrary to the Pyragas-type case, where the original resonance frequency contribution is enhanced, in the symmetric-cavity case the best squeezing occurs mostly off-resonance \cite{Nemet2016}. Another trend is that increasing time delay entails smaller offset frequency, as can be deduced from \refeq{eq:nu_c}.

We compare our results with the power spectra without feedback from \cite{Collett1987,Drummond1990} in \reffig{fig:sq_specs} for two different pump parameters, both of which are lower than the original threshold values. For smaller pump parameters, the feedback contributions are more dominant, as in that case the feedback rate is larger than the non-linearity. Notice the robustness of this enhancement as the obtained side-peaks can show stronger entanglement than the optimal value without feedback even for 5\% loss in the feedback loop (also presented in \reffig{fig:sq_specs}). 

\subsection{Stability analysis}\label{sec:Stability}

All of our calculations have been performed in the undepleted pump approximation below threshold, where the mean amplitudes of the resonator output fields are zero. Therefore, the previously obtained, frequency-dependent field operator solutions properly characterize the fluctuations around the steady state and can be used for linear stability analysis. Thus, rewriting the equations in terms of quadrature operators,
\begin{align}
	\frop{X}_\alpha&=\frac{\frop{\alpha}_{2,out}+\frop{\alpha}^\dagger_{2,out}}{\sqrt{2}},&
	\frop{Y}_\alpha&=\frac{\frop{\alpha}_{2,out}-\frop{\alpha}^\dagger_{2,out}}{i\sqrt{2}} ,
\end{align}
for $\alpha\in\lka a,b\rka$, we obtain the following equations for the stability eigenvalues (details in \refapp{app:stability}),
\begin{align}
	\label{eq:stab_a}
	\lsz R_a(\nu)+i\cdot I_a(\nu)\rsz\lsz R_b(\nu)+i\cdot I_b(\nu)\rsz-|\eps|^2&=0,\\
	\label{eq:stab_b}
	\lsz R_a(\nu)-i\cdot I_a(\nu)\rsz\lsz R_b(\nu)-i\cdot I_b(\nu)\rsz-|\eps|^2&=0 ,
\end{align}
where 
\begin{align}
	\label{eq:stab_a_det}
	R_a(\nu) &= \lambda+\kappa_a+k_a\cos{\phi_a}e^{-\lambda\tau_a},
	&I_a(\nu) &= \Delta+k_a\sin{\phi_a}e^{-\lambda\tau_a},\\
	\label{eq:stab_b_det}
	R_b(\nu) &= \lambda+\kappa_b+k_b\cos{\phi_b}e^{-\lambda\tau_b},
	&I_b(\nu) &= \Delta-k_b\sin{\phi_b}e^{-\lambda\tau_b}.
\end{align}
The curves corresponding to the real and imaginary parts of \refeq{eq:stab_a} and \refeq{eq:stab_b} are shown as insets within the spectral plots of \reffig{fig:sq_specs}. The intersections of real and imaginary parts --- indicated by red dots -- show the points where equations \refeq{eq:stab_a} and \refeq{eq:stab_b} are satisfied. 

The critical parameters are defined as the set where the eigenvalue with the largest real part crosses the imaginary axis. This means that its real part vanishes, triggering a stability change, or, in other words, a phase transition. Either parameter can be interpreted as an order parameter via which a critical point of the multidimensional stability landscape of the setup can be approached.

\section{Non-identical parameters for the two modes} \label{sec:NonIdenticalPars}
The results presented in the previous sections are based on the similarity in the descriptions of the NDPA and a combined system of two identical DPAs -- based on the matching parameters and the intrinsic entanglement between the down-converted modes. The obtained results justify these conclusions as the entanglement properties of the NDPA with TDCF in this regime show many similarities with the squeezing properties of the DPA \cite{Nemet2016}.
 
Considering non-degenerate modes, on the other hand, provides us with the unique opportunity to vary the parameters of the two down-converted modes independently, while investigating the induced changes in the entanglement properties. In the most general case, the critical frequency and time delay have a much more complicated expression than in \refeq{eq:nu_c}. The critical frequency corresponds to the imaginary part of the stability eigenvalue with the largest real part as it vanishes. Thus, all the parameters varying the stability landscape also have an influence on the entanglement properties of the setup. The appropriate expressions can be derived using the combination of \refeq{eq:stab_a} and \refeq{eq:stab_b}. 

This section aims to extend our investigations by departing from the doubled DPA picture. Considering different parameters in the two down-converted modes we study the influence of these differences in the feedback strengths $k_{a,b}$, the mode separation $\Delta$, feedback time delays $\tau_{a,b}$ and the feedback phases $\phi_{a,b}$ in detail to obtain further insights about the system's entanglement properties.

These investigations allows us to focus on the sensitivity of the current setup to these changes. Considering the two loops as two paths of excitations may be a step towards the application of the NDPA with TDCF as a quantum-enhanced interferometer.

\subsection{Difference in the feedback strengths} \label{sec:fb_strength_diff}

\begin{figure}[t!]
	\centering
	\vglue -.2 cm
	\includegraphics[width=\textwidth]{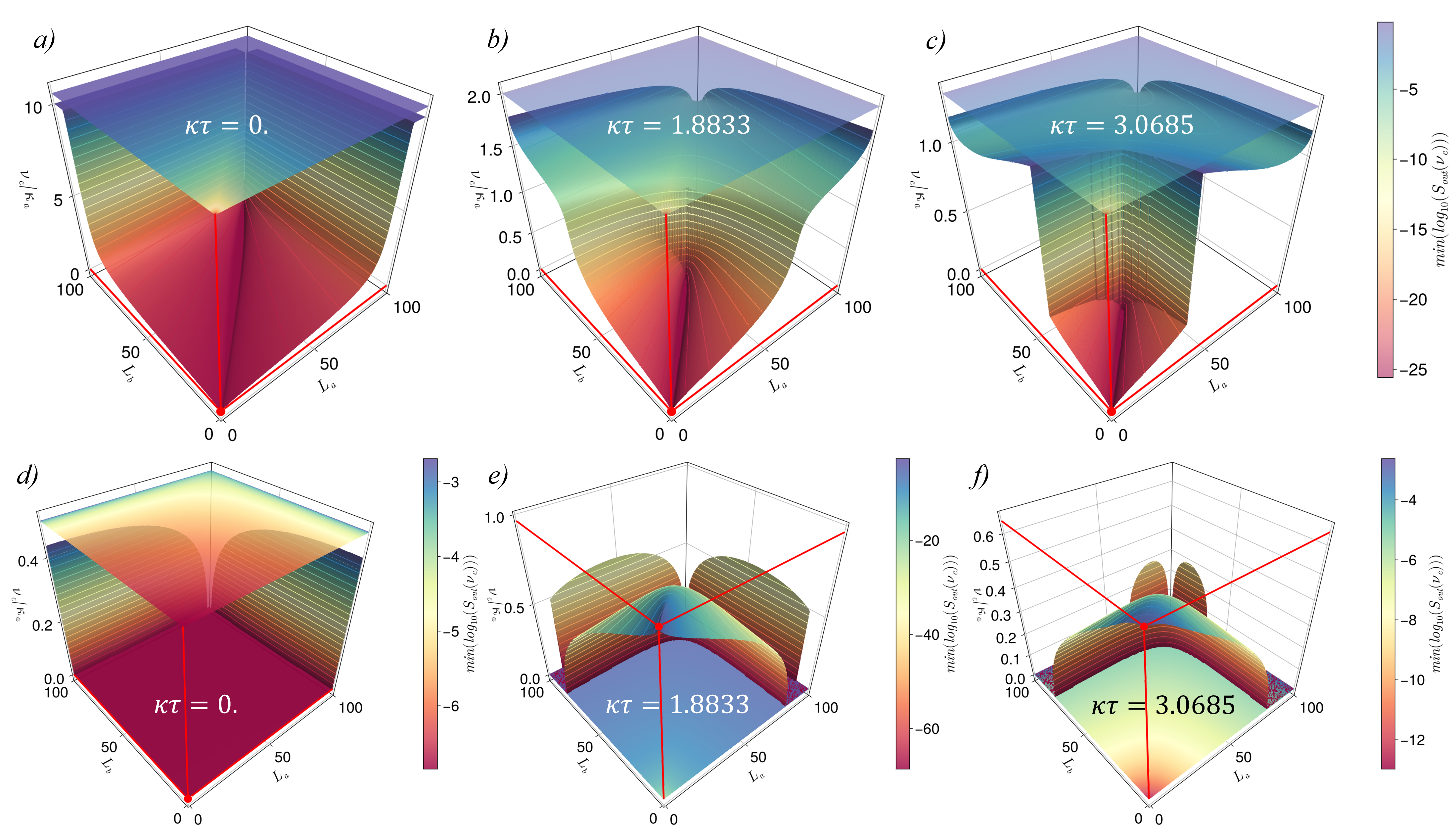}
	\vglue -.2 cm
	\caption{\linespread{1}The best available entanglement at certain loss rates in the feedback loops projected to a 2D surface at the bottom of the figure. The height represents the corresponding frequency as a function of the loss rates in the feedback loops.\lbCap Parameters: $\thetap=\pi,\Delta=0, \kappa_\alpha=10\cdot2\pi$ MHz, where $\alpha\in\lka a,b\rka$\lbCap \textit{a-c)} $\phi_{\alpha}=\pi, \varepsilon=0.45\kappa_a, \kappa_{1,\alpha}=0.9335\kappa_\alpha$, \textit{d-e)} $\phi_{\alpha}=0, \varepsilon=0.75\kappa_a, \kappa_{1,\alpha}=0.5\kappa_\alpha$
	}\label{fig:diffL}
	\vglue -.3 cm
\end{figure}

First, we focus specifically on the differences between the feedback strengths 
\begin{align} k_\alpha=2\sqrt{\kappa_{1,\alpha}\kappa_{2,\alpha}(1-L_\alpha)} \text{, where } \alpha\in\lka a,b\rka
\end{align}
which can be tuned in two ways in the present setup. One way is to vary the outcoupling rate of the feedback loop by tuning the reflectivity of the beam splitter - representing the loss rate - $L$ in the loop. Considering an almost one-sided and a symmetric resonator (see \reffig{fig:diffL}) a difference between these variables for the down-converted modes does not result in better entanglement. The best performance is obtained at $L_a=L_b=0$. Nevertheless, the addition of a long feedback delay introduces a degree of robustness around this solution.

\begin{figure}[b!]
	\centering
	\vglue -.2 cm
	\includegraphics[width=\textwidth]{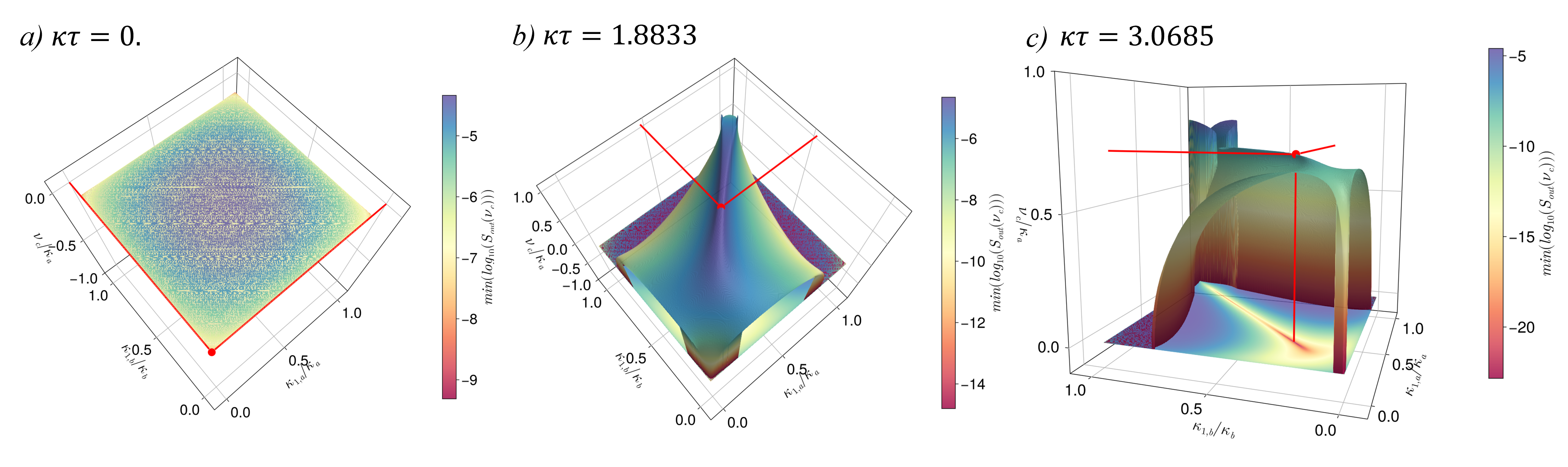}
	\vglue -.2 cm
	\caption{\linespread{1}The best available entanglement at certain outcoupling rates of the resonator mirrors for each mode projected to a 2D surface at the bottom of the figure. The height represents the corresponding frequency as a function of the outcoupling rate of the resonator mirrors for each mode.\lbCap Parameters: $\kappa_\alpha=10\cdot2\pi$ MHz, where $\alpha\in\lka a,b\rka,\thetap=\pi,\Delta=0, \phi_\alpha=0,L_{\alpha}=0,\epsilon=0.49\kappa_a$.
	}\label{fig:diffr}
\end{figure}

Tuning the feedback strength is also possible by altering the resonator outcoupling rates. (see \reffig{fig:diffr}). When no delay is considered, the best entanglement occurs for a truly one-sided resonator. However, the introduction of a substantial time delay results in strong entanglement even for varying reflectivities. The best performance is achievable for matching parameters in both down-converted modes which emphasizes the strong correlations between them.

A common feature in both cases (\reffig{fig:diffL} and \reffig{fig:diffr}) is that the feedback delay as a control parameter has a substantial influence on the degree and characteristic frequency of the strongest available entanglement. This is especially true for \reffig{fig:diffL} \textit{d-f)} and \reffig{fig:diffr}, where without time delay the resonance frequency dominates, while a substantial feedback delay changes this tendency by introducing a more complex stability landscape. Further examples can be seen in \refapp{app:tau_eff}.

\subsection{Difference in the Mode Frequencies}

\begin{figure}[b!]
	\centering
	\includegraphics[width=.9\textwidth]{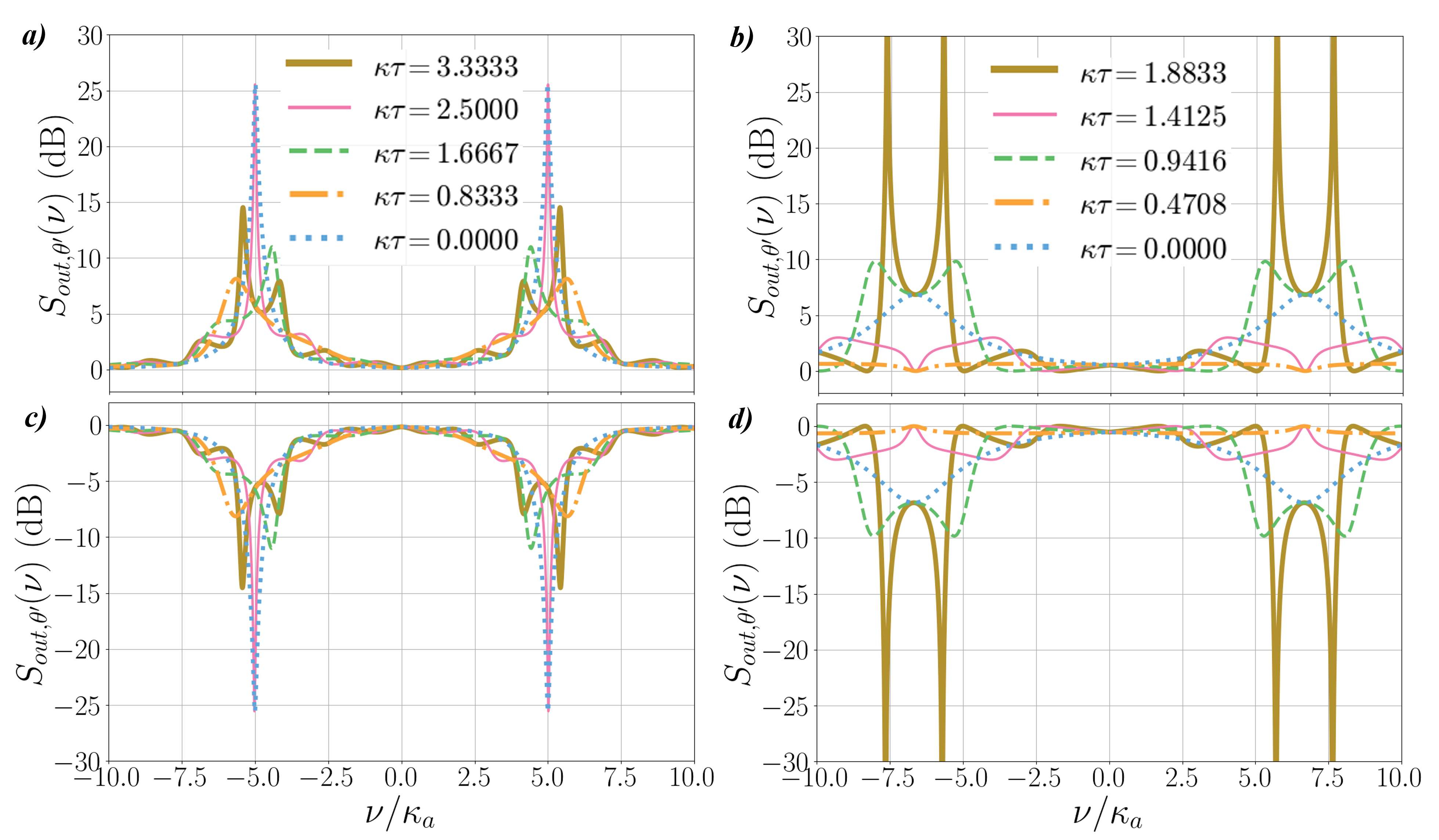}
	\vglue -0.3 cm
	\caption{\linespread{1}Noise power spectra in a) an almost one-sided and b) a symmetric cavity with coherent time-delayed lossless feedback.  \lbCap Parameters: $\kappa_\alpha=10\cdot2\pi$ MHz, where $\alpha\in\lka a,b\rka$ \lbCap
		left: $\phi_\alpha=\pi, |\eps|=0.45\kappa_a, \kappa_{1,\alpha} = 0.933\kappa_\alpha,\Delta=5\kappa_\alpha
		$, \lbCap
		right: $\phi_\alpha=0,|\eps|=0.75\kappa_a, \kappa_{1,\alpha} = \kappa_\alpha/2, \Delta = 4\pi\kappa_\alpha/1.8833. 
		$}\label{fig:det_delays}
	\vglue -.3 cm
\end{figure}

\noindent Having a finite frequency difference between the down-converted modes results in a divided two-mode noise power spectrum. The deviations from vacuum fluctuations are centered around the two down-converted mode frequencies, i.e., at $\pm\Delta$, which is shown in the frame rotating at half the pump frequency in \reffig{fig:det_delays}. As this frequency difference introduces extra phase shifts, new phase-matching conditions besides \refeq{eq:Delta_cond} are needed for the feedback,
\begin{align}
	\label{eq:detdel}
	\Delta\tau_{a,c}=\Delta\tau_{b,c}=n2\pi,\qquad n\in\mathbb{Z},
\end{align}
which translates as a condition for constructive interference at the input port between the two down-converted modes. This can also be seen from \refeq{eq:nu_c}. 

Note that the two sides of the entanglement spectrum centered around the down-converted frequencies are in general not the same; they are rather mirror images of each other due to the assumed relationship between the pump and down-converted modes. The reduced noise compared to the shot noise, signaled as dips in the spectrum, demonstrates the quantum nature of this correlation between the signal and idler fields as well.
Thus, focusing on a single down-converted mode may show an asymmetric spectrum, for example, in the Pyragas-type feedback case, which is similar to what was found in general for a parametric amplifier with a pump detuned from the cavity resonance \cite{Carmichael1984}. However, due to energy-conservation considerations, the overall spectrum should be symmetric.

\begin{figure}[b!]
	\centering
	\includegraphics[width=.95\textwidth]{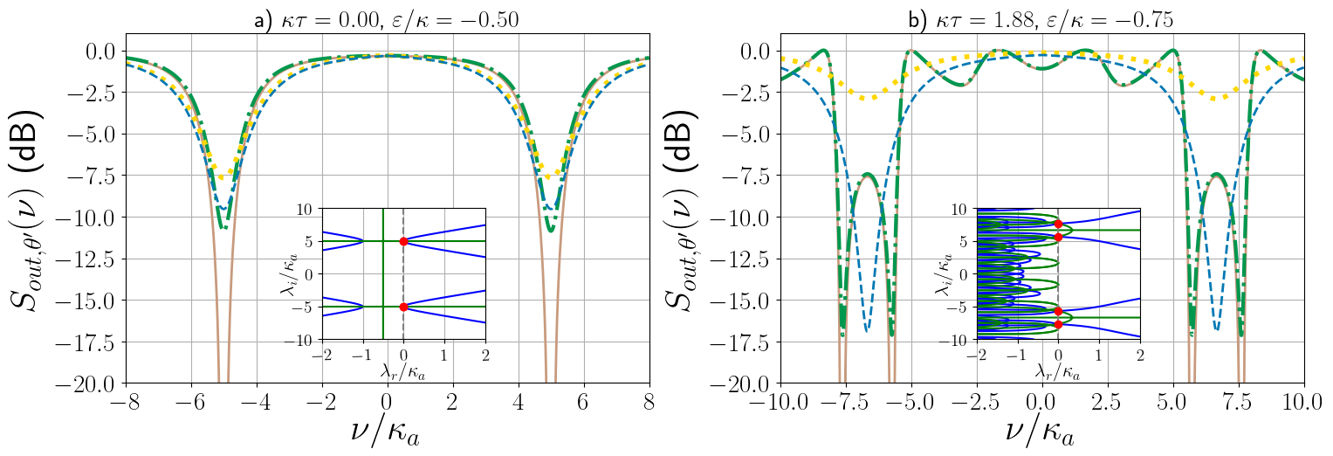}
	\vglue -0.4 cm
	\caption{{\linespread{1}Noise power spectra without feedback in a symmetric ($\kappa_{1,\alpha}=\kappa_{2,\alpha}, \alpha\in\{a,b\}$ yellow dotted) and one-sided cavity ($\kappa_{1,\alpha}=\kappa_{\alpha}$ blue dashed) and in a symmetric cavity with a lossy (green dash-dotted) and lossless feedback loop (brown solid). The inset shows the solution curves for the stability equations \refeq{eq:stab_a} and \refeq{eq:stab_b} with red dots signaling the eigenvalues with the largest real part.\lbCap Parameters: $\thetap=\pi$, $L_\alpha=5\%$ (for the lossy TDCF case),  $\kappa_\alpha=\kappa$, $\tau_\alpha=\tau$, a) $|\eps|=\kappa/2$, $\phi_\alpha=0$, $\Delta=5\kappa$, $\tau=0$, b) $|\eps|=0.75\kappa$, $\phi_\alpha=0$, $\Delta=6.6726\kappa$, $\kappa\tau=1.8833$.}}\label{fig:det_sq_specs}
	\vglue -.1 cm
\end{figure}

Note that for Pyragas-type feedback the feedback phase is $\pi$, which means that we would expect only a single peak for one half of the spectrum as in \reffig{fig:delays} $a)$ and $c)$. In \reffig{fig:det_delays} this is only true for the blue dotted and pink solid curves, as these delay values are the ones that are close enough to satisfy the characteristic equation (\ref{eq:detdel}). As we get further away from this condition in parameter space, the form of the noise power spectrum also deviates, i.e., it is divided around the detuned resonance. Stability considerations based on \refeq{eq:stab_a_det} and \refeq{eq:stab_b_det} also show that for a phase shift of $\phi_\alpha=\pi \lk\alpha\in\lka a,b\rka\rk$ in the feedback loop, feedback plays a role mostly in the position of the peaks. This shows that, for a Pyragas-type time-delayed feedback, the above defined phase-matching condition becomes a dominant factor in determining the shape of the output field noise spectrum.

In the phase-matched feedback case, matching of the frequency-shifted and delay-induced as well as the intrinsic feedback phase have to be satisfied. This can also be seen in \reffig{fig:det_delays}, where in $b)$ and $d)$ the competition between the two cancels the peaks clearly visible in \reffig{fig:delays} (orange dash-dotted and pink solid curves) at the mode-resonance. In other words, based on the stability considerations of \refeq{eq:stab_a_det} and \refeq{eq:stab_b_det}, for a phase shift of $\phi_\alpha=0 \lk\alpha\in\lka a,b\rka\rk$ in the feedback loop, the feedback parameters have a significant influence not only on the position but on the size of these peaks, as well.

\begin{figure}[b!]
	\centering
	\vglue -.2 cm
	\includegraphics[width=.95\textwidth]{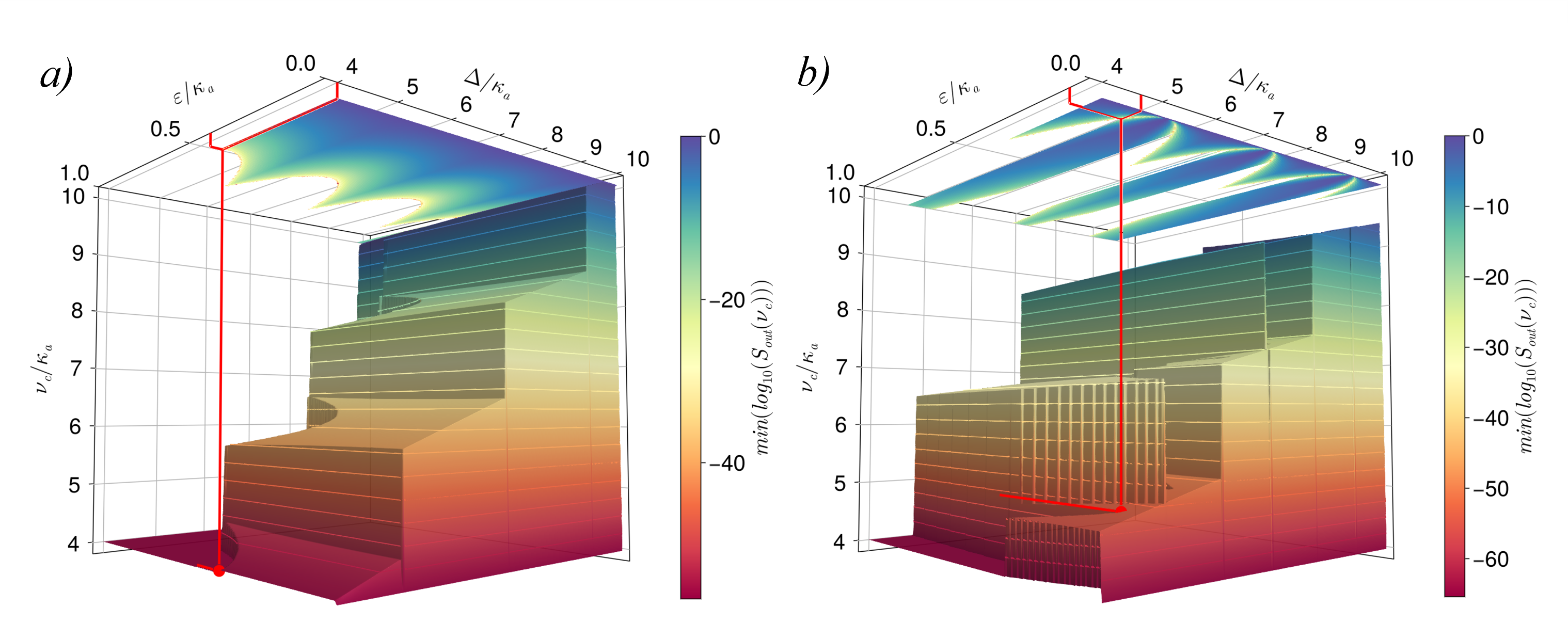}
	\vglue -.2 cm
	\caption{\linespread{1}The height shows the frequency corresponding to the strongest entanglement, whereas the heatmap at the top shows the measure of the lowest noise power as a function of the frequency difference between the down-converted modes $\Delta$ and the pump parameter $|\epsilon|$. The red dot signals the minimal noise, the best entanglement that can be obtained within the investigated parameter range. The auxiliary red lines point to the actual parameter values and the corresponding characteristic frequency.\lbCap
		Parameters: $\thetap=\pi,\kappa_\alpha=10\cdot2\pi$ MHz, where $\alpha\in\lka a,b\rka$, $\kappa_\alpha\tau_\alpha=3.0865,L_\alpha=0\%$, a)$\kappa_{1,\alpha}=0.933\kappa_\alpha,\phi_\alpha=\pi$, b) $\kappa_{1,\alpha}=\kappa_\alpha/2,\phi_\alpha=0$. 
	}\label{fig:eps-del_map}
	\vglue -.3 cm
\end{figure}

Nevertheless, with all phase-matching and resonance conditions satisfied, \reffig{fig:det_sq_specs} shows that similar enhancement of entanglement can be achieved as for the DPA. An interesting and important characteristic of \refeq{eq:nu_c} is that it allows for different values of $\Delta$ as well, provided that the mirror transmission rates and/or the pump parameter are adjusted, as can be seen in \reffig{fig:eps-del_map}. We note that such flexibility is feasible, e.g., with parametric amplifiers in superconducting circuits  \cite{Castellanos-Beltran2007}, or in a nanofiber-based scheme \cite{Yeo2013,Zlobina2013}. In this case only the following equation should hold, based on
\refeq{eq:stab_a} and \refeq{eq:stab_b}:
\begin{align}
	\label{eq:gen_cond}
	\cos{(\Delta\tau_{a,c})}\cos{\lsz\sqrt{k_a^2-(|\eps|-\kappa_a)^2}\tau_{a,c}\rsz}\pm
	\sin{(\Delta\tau_{a,c})}\sin{\lsz\sqrt{k_a^2-(|\eps|-\kappa_a)^2}\tau_{a,c}\rsz}=
	|\eps|-\kappa_a .
\end{align}
The solutions for this equation also represent the poles of the system in the frequency domain. Therefore, by changing the frequency difference between the modes, the stability oscillates between stable and unstable regimes, as can also be seen in \reffig{fig:eps-del_map}. The values of the characteristic frequencies are given by
\begin{align}
	\nu_c = \pm\lk\Delta\pm\sqrt{k_a^2-(|\eps|-\kappa_a)^2}\rk ,
\end{align}
where dips in the two-mode squeezing spectrum can be observed.

The present scheme is also robust as it shows a level of resistance to feedback losses. Although with the introduction of loss in the feedback loop, the sharp resonant character is lost, the prepared entanglement is still stronger than originally. In the example depicted in \reffig{fig:eps-del_map}, 20 dB noise reduction is obtained in the general quadrature even with 5\% loss, which is a feasible range for both circuit-QED and optical fibre setups.

\subsection{Difference in the time delays}

In terms of dynamical time scales, time delays in the feedback loops play a fundamental role.
Following the same steps as in \cite{Nemet2016}, we look at the dynamics using a semiclassical mean-field approximation, where depletion of the pump field amplitude is included and its equation of motion is also added.
\begin{align}
	\tder{\alpha} &= -(\kappa_a+i\Delta)\alpha(t) + \chi\beta^*(t)\gamma(t)-k_ae^{i\phi_a}\alpha(t-\tau_a),\\
	\tder{\beta} &= -(\kappa_b-i\Delta)\beta(t) + \chi\alpha^*(t)\gamma(t)-k_be^{i\phi_b}\beta(t-\tau_b),\\
	\tder{\gamma} &= -\kappa_\gamma\gamma(t) - \chi\alpha(t)\beta(t)+\eta,
\end{align} 
where $\alpha,\beta,\gamma$ stand for the complex, semi-classical amplitudes of $\aop,\bop$, and the pump field, respectively, $\kappa_\gamma$ is the damping rate of the pump field amplitude, $\chi$ denotes the coupling between the pump and the down-converted modes (i.e., characterizes the second-order non-linearity), and $\eta$ is the complex coherent driving amplitude of the pump field, such that
$\varepsilon = (\eta /\kappa_\gamma )\chi$.

\begin{figure}[t!]
	\centering
	\vglue -.2 cm
	\includegraphics[width=.85\textwidth]{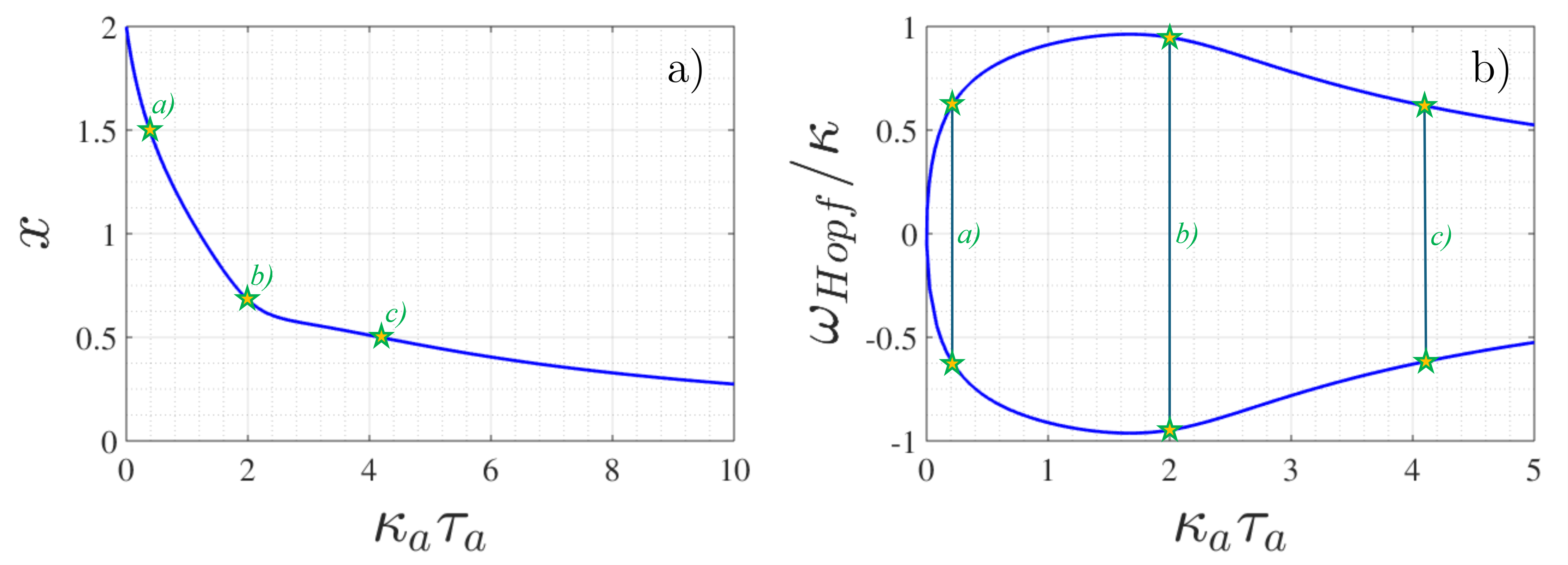}
	\vglue -.2 cm
	\caption{\linespread{1}a) Continuation of the Hopf-bifurcation curve for varying $x=\eta/\kappa_a$ and $\kappa_a\tau_a$, while fixing $\kappa_b\tau_b$. The curve indicates the critical parameter values where stability change of the below threshold steady states occur. For large enough $x$, oscillating solutions emerge, which can be obtained using DDE-BIFTOOL. b) The characteristic oscillation frequencies are presented as a function of $\kappa_a\tau_a$. Example spectra are plotted in \reffig{fig:diff_delay} at the parameter values signaled by stars.\lbCap
		Parameters: $\Delta=0,\kappa_\alpha=10\cdot2\pi$ MHz, where $\alpha\in\lka a,b\rka$, $\kappa_{1,a}=\kappa_{1,b}=\kappa_a/2, \kappa_b\tau_b = 2,\phi_a=\phi_b=0$. }\label{fig:Hopf}
\end{figure}

\begin{figure}[b!]
	\centering
	\vglue -.2 cm
	\includegraphics[width=.99\textwidth]{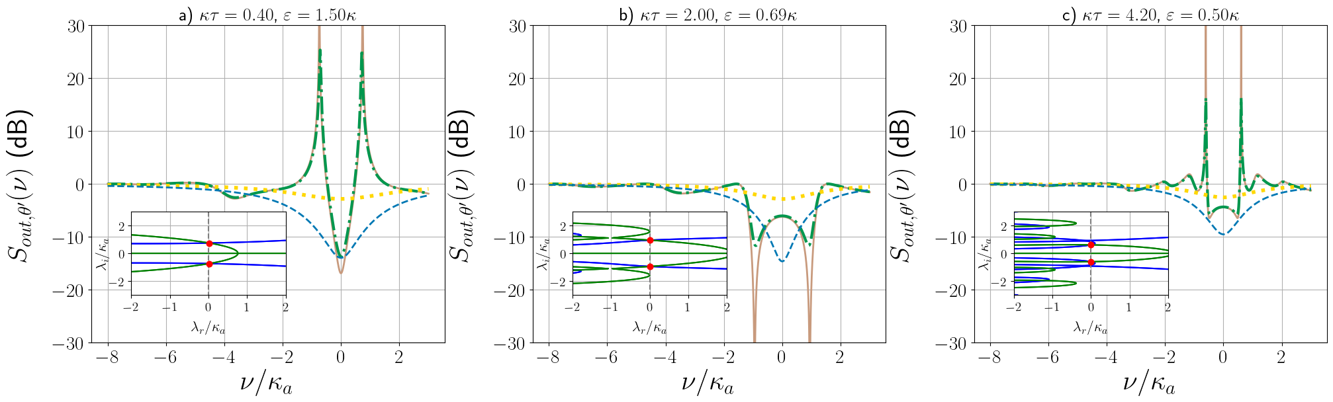}
	\vglue -.2 cm
	\caption{\linespread{1} Noise power spectra for the parameter combinations at the three points on the curves of \reffig{fig:Hopf}. For comparison we also plot the case without feedback in a symmetric ($\kappa_{1,\alpha}=\kappa_{2,\alpha}, \alpha\in\{a,b\}$ yellow dotted) and one-sided cavity ($\kappa_{1,\alpha}=\kappa_{\alpha}$ blue dashed) and in a symmetric cavity with a lossy (green dash-dotted) and lossless feedback loop (brown solid). The inset shows the solution curves for the stability equations, with red dots signaling the eigenvalues with the largest real part.\lbCap
		Parameters: $\thetap=\pi$, $\Delta=0$, $\kappa_\alpha=10\cdot2\pi$ MHz, $\kappa_{1,\alpha}=\kappa_\alpha/2$, $\kappa_b\tau_b = 2$, $\phi_\alpha=0$, $L_\alpha=5\%$ (for the lossy TDCF case). }\label{fig:diff_delay}
	\vglue -.3 cm
\end{figure}

In order to examine how difference in the feedback delays affect the stability landscape, we start from a critical point showing Hopf bifurcation - or persistent oscillations - in the parameter space of driving strength $(x)$ and time delay $(\kappa_a\tau_a)$ in one mode with no mode separation $(\Delta=0)$. Then a continuation curve is drawn by following the criticality while varying the driving strength or the time delay in the other mode (see \reffig{fig:Hopf}a)). As the driving strength is increased, changing stability for the below-threshold results is expected at shorter time delays. Meanwhile, for longer time delays, lower pump parameter values suffice to induce oscillations.

The imaginary part of the critical stability eigenvalues are presented in \reffig{fig:Hopf}b), which show the oscillating character of the emerging periodic solutions. Note that oscillations appear only when the time delay of the feedback loop is sufficiently large compared to the inverse decay rate of the resonator. Increasing the length of the feedback loop, at first the characteristic frequency of Hopf bifurcation - or observed oscillations - shifts away from resonance, then, upon reaching the cavity linewidth, it starts to decrease again. 

\begin{figure}[b!]
	\centering
	\vglue -.2 cm
	\includegraphics[width=.65\textwidth]{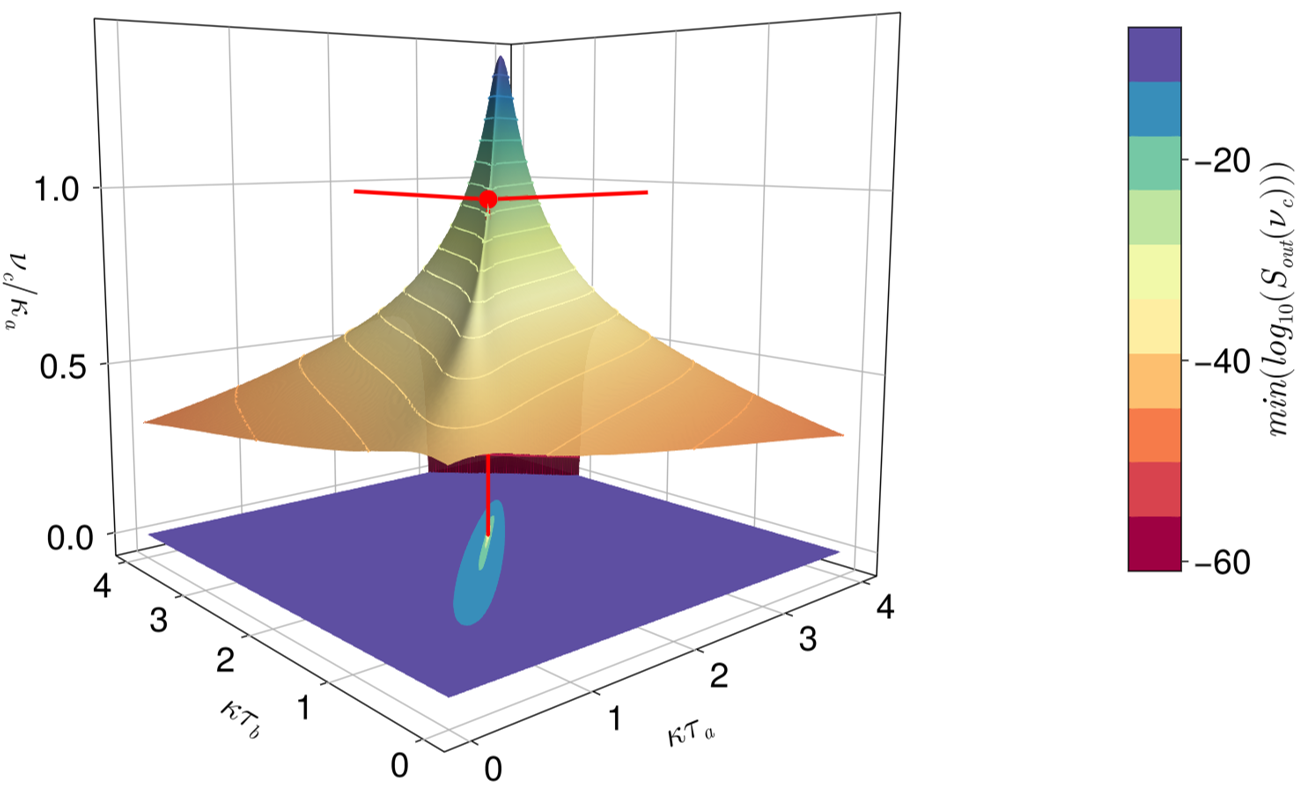}
	\vglue -.2 cm
	\caption{\linespread{1}The best available entanglement as a projection to the bottom surface together with the corresponding critical frequency as a function of the time delays in the feedback loops.\lbCap
		Parameters: $\thetap=\pi,\Delta=0,\kappa_\alpha=10\cdot2\pi$ MHz, where $\alpha\in\lka a,b\rka$, $\kappa_{1,\alpha}=\kappa_\alpha/2,\eps=3\kappa_\alpha/4, \phi_\alpha=0,L_\alpha=0\%$. }\label{fig:Smin_nuc}
	\vglue -.3 cm
\end{figure}

Investigating the strength of entanglement at three particular points on the curves in \reffig{fig:Hopf} gives the corresponding noise power spectra in \reffig{fig:diff_delay}. \reffig{fig:Hopf}a) together with \reffig{fig:diff_delay} convey an important message about the influence of TDCF on the quantum properties of the system. Even though the stability changes appear as singularities in the noise power spectra, only the case of matching time delays in the two modes result in reduced fluctuations instead of antisqueezing.
The same can be seen in \reffig{fig:Smin_nuc}, where the frequencies of the strongest entanglement are plotted with the corresponding noise power $S_{out}$. These observations further emphasize the sensitivity of the present scheme as the chosen parameter combinations for strong entanglement are a result of finely tuned interference.

Introducing a finite loss in the feedback loop slightly shifts the position of the strongest entanglement in frequency and regulates its singular character. Nevertheless, the noise reduction at those frequencies can still be greater than in the original setup. In the case of the example depicted in \reffig{fig:diff_delay}, even 15 dB noise reduction is achievable with a symmetric cavity with 5\% feedback loss, which is a feasible range for both circuit-QED and optical fiber setups.

\subsection{Difference in the phase shifts of the loops}

\begin{figure}[b!]
	\centering
	\vglue -.2 cm
	\includegraphics[width=.85\textwidth]{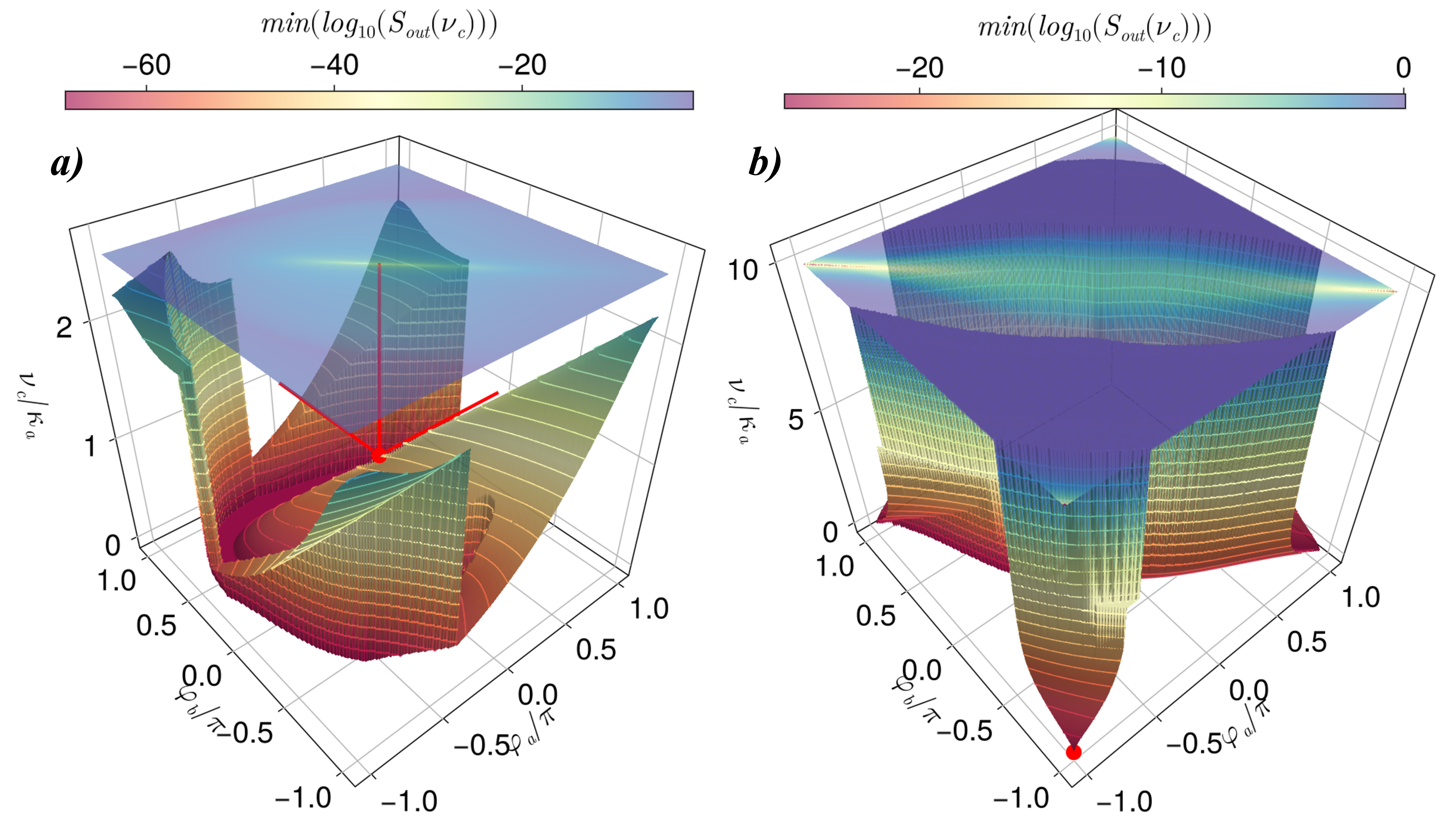}
	\vglue -.2 cm
	\caption{{\linespread{1}The best available entanglement at certain phase values in the feedback loops projected to a 2D surface at the top of the figure. The height represents the corresponding critical frequency as a function of the phase shifts in the feedback loops. \lbCap Parameters: $\thetap=\pi,\Delta=0, L_a=L_b=0\%,\kappa_\alpha=10\cdot2\pi$ MHz, where $\alpha\in\lka a,b\rka$,\lbCap
			a) $\eps=3\kappa_\alpha/4, \kappa_{1,\alpha}=\kappa_\alpha/2, \kappa\tau_\alpha=1.8833$.
			b) $\eps=0.45\kappa_\alpha,  \kappa_{1,\alpha}=0.933\kappa_\alpha, \kappa\tau_\alpha=0$,
	}}\label{fig:diff_phase}
\end{figure}

In this last subsection, we investigate how differences in the feedback phase shifts influence the obtained noise characteristics. As quantum interference plays a crucial role in the properties of the output fields of the system, the highest entanglement is only achievable when the condition
\begin{align}
	\phi_a+\phi_b= n2\pi
\end{align}
is met. Its strength can still be substantial in non-ideal regimes as well, as demonstrated in \reffig{fig:diff_phase} for both a symmetric and an almost one-sided resonator with TDCF. The range of parameter space for strong entanglement is larger for a one-sided cavity. In that case, the best performance is achieved on resonance when $\phi_a=\phi_b=\pi$. In the symmetric-cavity case, the strongest entanglement is obtained with complete phase-matching of $\phi_a=\phi_b=0$ at a finely tuned characteristic frequency set by the parameters of the setup as in \refeq{eq:gen_cond}.

Measurement-based feedback techniques have been applied for phase-estimation with a quantum-enhanced sensitivity \cite{Berni2015a,Clark2016b}. The above mentioned characteristics of the proposed setup also suggest a high potential for its application in the field of quantum enhanced interferometry, however this falls outside the scope of the present work. 
\section{Summary}

This paper presents a detailed study about the variations TDCF may induce in the EPR-type entanglement properties of an NDPO. The calculations are performed in the undepleted pump regime with no seeds for the down-converted modes and the pump field considered to be constant and coherent. The two-mode squeezing spectrum of a generalized quadrature is used as a figure of merit which combines the noise contributions from both modes.

The extension of coherent feedback loops introduce additional control parameters that can tune the frequency, bandwidth, and the strength of the best available entanglement. Most of these quantities, such as the time delay, phase, and loss of the feedback loop might be adjusted experimentally with the addition of phase shifters and tunable beam splitters. 
The emerging strong entanglement is identified as a combination of changing stability landscape and a unique interplay of quantum fluctuations and system coherence between the past and present fields at the point of contact. 

The ever-growing attention drawn to quantum networks motivate experimental conditions to develop at a fast pace, thus the prospect of experimental study of longer coherent feedback extensions seems promising. The intrinsic robustness of the presented scheme is advantageous in this process, as substantial enhancement of entanglement is possible compared to the case without feedback, even with finite loss in the feedback loop. These ``non-ideal'' regimes might actually be better for operation, as in this case the stability of the performance allows for a wider range of tunability in the parameters.

The last part, \refsec{sec:NonIdenticalPars} of the paper 
focuses on the unique potential of the NDPO setup to observe the effects of differences in feedback parameters between the two modes. As the entanglement properties are strongly tied to the semiclassical dynamics of the system, the introduction of time delay brings in unique possibilities to control the entanglement characteristics. With the introduction of TDCF one can, for example, counteract the random frequency shifts of the down-converted modes by adjusting the time delay in the feedback loop. The required pump intensity for the same entanglement strength can also be tuned by varying the resonator outcoupling rate and the time delay. 

Finally, the last section also shows the setup's intrinsic sensitivity to phase variations, which makes this arrangement a promising platform for precision measurement applications, for example in an SU(1,1) interferometric setup. This idea provides a good basis for future work, but falls outside the scope of the present manuscript. Investigations could also be extended to cover the non-linear, depleted pump regime as well.
\section*{Acknowledgements}

This research was supported
by the Ministry of Culture and Innovation, and  the National Research, Development and Innovation Office within the Quantum Information National Laboratory of Hungary (Grant No. 2022-2.1.1-NL-2022-00004) and the project Grant No. TKP-2021-NVA-04. The contribution of S.P. was supported in part by grant NSF PHY-2309135 to the Kavli Institute for Theoretical Physics (KITP).

\clearpage
\appendix
\section{Time evolution derivation}\label{app:Tev_der} 
\noindent The most general form of the time evolution of a linear open quantum system in the Heisenberg picture can be written as  \cite{Petruccione2007,Whalen2016}:
\begin{align}
	\label{eq:gen_Heis}
	\frac{d}{dt}\hat{\underline{a}}(t)& = \frac{i}{\hbar}\lsz\Hop_S, \hat{\underline{a}}(t)\rsz - 
	\int_0^t\dul{f}(t-t^\prime)\hat{\underline{a}}(t^\prime) - 
	i\underline{\Omega}(t)-\hat{\underline{B}}(t),\\
	\underline{\aop}(t) &= \lsz \aop(t),\adop(t),\bop(t),\bdop(t)\rsz,\\
	\underline{\Omega}(t)&= 0,\\
	\underline{\Bop}(t)&= \lsz\sqrt{2\kappa_a}\aop_{in}(t),\sqrt{2\kappa_a}\adop_{in}(t),\sqrt{2\kappa_b}\bop_{in}(t),\sqrt{2\kappa_b}\bdop_{in}(t)\rsz,
\end{align}
where $\underline{\aop}$ is the collection of system operators (in our case, operators for the two modes of the NDPO) and $\Hop_S$ represents the Hamiltonian that guides the closed-system dynamics. The imprint of the environmental structure -- whether it is Markovian or non-Markovian, for example -- shapes the system dynamics as well. This imprint is gathered in the so-called dissipation kernel $\dul{f}(t)$. External driving fields can be added as $\underline{\Omega}(t)$. Finally, the characteristic quantum noise 
describing the intrinsic uncertainties of quantum fields in the setup is added as stochastic input field operators, $\underline{\Bop}(t)$. 

The finite time delay of the feedback in our system translates into non-Markovian evolution, corresponding to introducing a memory into the system dynamics. The dissipation kernel provides a common description for both Markovian and non-Markovian reservoir structures. It is related to the quantum noise operators ($\underline{\Bop}(t)$) via the spectral density $\dul{J}_{\alpha}(\omega)$ of the environments (labelled with $\alpha$) that are interacting separately with the system:
\begin{align}
	\label{eq:FJ}
	f_{jk}(t_{1}-t_{2})=\lsz\Bop_j(t_1),\Bdop_k(t_2)\rsz=\sum_\alpha\int_{-\infty}^\infty J_{\alpha jk}(\omega)e^{-i\omega (t_{1}-t_{2})}d\omega,
\end{align}
where an extra minus sign is needed when $\Bop_j(t)$ is a creation operator. Markovian interactions associated with the loss channels in our system correspond to
\begin{align}
	\label{eq:Ja}
	J_{\xi_ajk}(\omega) &= \frac{\kappa_{2,a}L_a}{\pi}\delta_{jk}, & j,k&\in\{1,2\},\\
	\label{eq:Jb}
	J_{\xi_bjk}(\omega) &= \frac{\kappa_{2,b}L_b}{\pi}\delta_{jk}, &  j,k&\in\{3,4\}.
\end{align}
The non-Markovian feedback loops - with finite time-delays $\tau_{\alpha,\beta}$ and losses in the respective loops $L_\alpha,L_\beta$ - are represented by
\begin{align}
	\label{eq:Ja_d}
	J_{a_{1,in}jk}(\omega) &= \lsz\frac{\kappa_{1,a}+\kappa_{2,a}(1-L_a)}{\pi} + \frac{k_a}{\pi}\cos{(\omega\tau_a+\phi_a)}\rsz\delta_{jk},  & j,k&\in\{1,2\},\\
	\label{eq:Jb_d}
	J_{b_{1,in}jk}(\omega) &= \lsz\frac{\kappa_{1,b}+\kappa_{2,b}(1-L_b)}{\pi} + \frac{k_b}{\pi}\cos{(\omega\tau_b+\phi_b)}\rsz\delta_{jk}, &  j,k&\in\{3,4\},
\end{align}
where
\begin{align}
	k_\alpha &= 2\sqrt{\kappa_{1,\alpha}\kappa_{2,\alpha}(1-L_\alpha )} , \quad \alpha\in\{a,b\} ,
\end{align}
is the feedback strength.
Note that, for the distinguishability of environmental modes in \refeq{eq:Ja} and \refeq{eq:Jb}, non-overlapping feedback loops or orthogonal polarizations are considered.
\section{Stability analysis}\label{app:stability}
\noindent In the paper about DPA with TDCF \cite{Nemet2016}, it was shown that the highest levels of squeezing correspond to changing stability. Phase matching was an important aspect there as well, however, the degenerate case was more simple in a sense that all stability changes from the non-amplified regime were accompanied by singular noise reduction in one quadrature and  runaway noise in the other. Therefore, stability considerations of the system dynamics determined the validity of the linear model, experimental feasibility, and identification of enhanced entanglement. The expectation values of equations of motion \refeq{eq:aop_der} and \refeq{eq:bop_der} can be rearranged in the following way:
\begin{align*}
	\frac{d}{dt}\ul{X} &= -\dul{\Delta}_X\ul{X}(t)-\dul{\Phi}\,\ul{X}_\tau(t),\\
	\ul{X}(t) &= \lka X_{a,\thetap_a}(t),Y_{a,\thetap_a}(t),X_{b,\thetap_b}(t),Y_{b,\thetap_b}(t)\rka,\\
	\ul{X}_\tau(t) &= \lka X_{a,\thetap_a}(t-\tau_a),Y_{a,\thetap_a}(t-\tau_a),X_{b,\thetap_b}(t-\tau_b),Y_{b,\thetap_b}(t-\tau_b)\rka,\\
	\dul{\Delta}_X &= \left(\begin{matrix} 
		\kappa_a & -\Delta & -|\epsilon| & 0 \\
		\Delta & \kappa_a & 0 & |\epsilon| \\
		-|\epsilon| & 0 & \kappa_b & \Delta \\
		0 & |\epsilon| & -\Delta & \kappa_b\end{matrix}\right), \quad \dul{\Phi} =\left(\begin{matrix} 
		k_a\cos{\phi_a} & -k_a\sin{\phi_a} & 0 & 0 \\
		k_a\sin{\phi_a} & k_a\cos{\phi_a} & 0 & 0 \\
		0 & 0 & k_b\cos{\phi_b} & -k_b\sin{\phi_b} \\
		0 & 0 & k_b\sin{\phi_b} & k_b\cos{\phi_b}\end{matrix}\right).
\end{align*}

In order to determine whether a steady state is stable or not, its vicinity is described by exponential time-evolution $(X_i(t) = X_i(0)e^{\lambda t},Y_i(t) = Y_i(0)e^{\lambda t})$. Thus, the following should be satisfied:
\begin{align*}
	0 &=\left(\begin{matrix} 
		R_a(\nu) & -I_a(\nu) & -|\eps| & 0 \\
		I_a(\nu) & R_a(\nu) & 0 & |\eps| \\
		-|\eps| & 0 & R_b(\nu) & I_b(\nu) \\
		0 & |\eps| & -I_b(\nu) & R_b(\nu)\end{matrix}\right)\left(\begin{matrix}\tilde{X}_a\\\tilde{Y}_a\\\tilde{X}_b\\\tilde{Y}_b \end{matrix}\right),
\end{align*}
where
\begin{align*}
	R_a(\nu) &= \lambda+\kappa_a+k_a\cos{\phi_a}e^{-\lambda\tau_a},
	&I_a(\nu) &= \Delta+k_a\sin{\phi_a}e^{-\lambda\tau_a},\\
	R_b(\nu) &= \lambda+\kappa_b+k_b\cos{\phi_b}e^{-\lambda\tau_b},
	&I_b(\nu) &= \Delta-k_b\sin{\phi_b}e^{-\lambda\tau_b},
\end{align*}
which simplifies to:
\begin{align}
	\label{eq:stab1}
	\lsz R_a(\nu)+i\cdot I_a(\nu)\rsz\lsz R_b(\nu)+i\cdot I_b(\nu)\rsz-|\eps|^2&=0,\\
	\label{eq:stab2}
	\lsz R_a(\nu)-i\cdot I_a(\nu)\rsz\lsz R_b(\nu)-i\cdot I_b(\nu)\rsz-|\eps|^2&=0.
\end{align}
The numerical solution curves for these two equations are plotted as insets of the noise power spectrum figures. Red dots show their intersections with the largest real part, i.e., the eigenvalues determining the overall stability of the system.
\section{Short overview of two-mode squeezing noise spectrum}\label{app:2ModeSqueez}

\noindent Let us define the quadratures of the two down-converted modes $\aop$ and $\bop$ as
\begin{align}
	\hat{X}_{a,\thetap_a}&=\frac{1}{\sqrt{2}}\lk\aop e^{-i\thetap_a/2}+\adop e^{i\thetap_a/2}\rk,&\hat{X}_{b,\thetap_b}&=\frac{1}{\sqrt{2}}\lk\bop e^{-i\thetap_b/2}+\bdop e^{i\thetap_b/2}\rk,\\
	\hat{Y}_{a,\thetap_a}&=\frac{1}{i\sqrt{2}}\lk\aop e^{-i\thetap_a/2}-\adop e^{i\thetap_a/2}\rk,&
	\hat{Y}_{b,\thetap_b}&=\frac{1}{i\sqrt{2}}\lk\bop e^{-i\thetap_b/2}-\bdop e^{i\thetap_b/2}\rk.
\end{align}
Using these quadratures, a general quadrature can be defined in the following way:
\begin{align}
	\hat{X}_\thetap^G&=\hat{X}_{a,\thetap_a}+\hat{X}_{b,\thetap_b},&
	\label{eq:YG_def}
	\hat{Y}_\thetap^G&=\hat{Y}_{a,\thetap_a}-\hat{Y}_{b,\thetap_b},
\end{align}
where $
\thetap=\theta-(\thetap_a+\thetap_b)/2$.
Two-mode squeezing is identified using the sum of the normal-ordered variances of $\hat{X}_\thetap^G$ and $\hat{Y}_\thetap^G$ via the inequality \cite{Collett1987,Drummond1990,Caves1985}
\begin{align}
	\label{eq:var}
	\left\langle:\lk\Delta\hat{X}_\thetap^G\rk^2:\right\rangle+\left\langle:\lk\Delta\hat{Y}_\thetap^G\rk^2:\right\rangle<0,
\end{align}
where $\left\langle\lk\Delta\hat{O}\rk^2\right\rangle=\left\langle\hat{O}^2\right\rangle-\left\langle\hat{O}\right\rangle^2$.
From now on we suppose that $\theta=0$, thus $\thetap = -(\thetap_a+\thetap_b)/2$. For a heterodyne measurement setup $\thetap_a=\thetap_b$, whereas generally in a combined homodyne detection the local oscillator phases can differ.

Throughout this paper we use the noise power spectrum to describe the entanglement properties of the system,
\begin{align}
	S_{out,\thetap}(\nu) &= 10\log_{10}{\lsz\frac{\chi_{\thetap,out}(\nu)}{\chi_{SN}(\nu)}\rsz},\\
	\chi_{\thetap,out}(\nu)&=\int_{-\infty}^\infty\lsz\left\langle\lk\Delta\hat{\tilde{X}}_{\thetap,out}^G\rk^2(\nu,\nup)\right\rangle+\left\langle\lk\Delta\hat{\tilde{Y}}_{\thetap,out}^G\rk^2(\nu,\nup)\right\rangle\rsz d\nup\nn\\&=1+\int_{-\infty}^\infty\chi_N(\nu,\nup)d\nup ,
	\label{eq:chi_nu}
\end{align}
where $\chi_{SN}(\nu)$ denotes the shot-noise (vacuum) two-mode quadrature variance and $\chi_{N}(\nu,\nup)$ corresponds the "normal-ordered" variance:
\begin{align}
	\chi_{N}(\nu,\nup)&=\frac{1}{2}\lka\av{\awdop_{2,out}(-\nu)\awop_{2,out}(\nup)}+\av{\awdop_{2,out}(-\nup)\awop_{2,out}(\nu)}+\av{\bwdop_{2,out}(-\nu)\bwop_{2,out}(\nup)}\right.\nn\\&\left.+\hspace{.3cm}
	\av{\bwdop_{2,out}(-\nup)\bwop_{2,out}(\nu)}+\av{\awop_{2,out}(\nu)\bwop_{2,out}(\nup)}+\av{\bwop_{2,out}(\nu)\awop_{2,out}(\nup)}\right.\nn\\&\left.\hspace{.3cm}+\av{\awdop_{2,out}(-\nu)\bwdop_{2,out}(-\nup)}+\av{\bwdop_{2,out}(-\nu)\awdop_{2,out}(-\nup)}\rka
\end{align}
\section{Introduction of time delay for non-degenerate modes}\label{app:tau_eff}

\begin{figure}[h!]
	\centering
	\vglue -.2 cm
	\includegraphics[width=.95\textwidth]{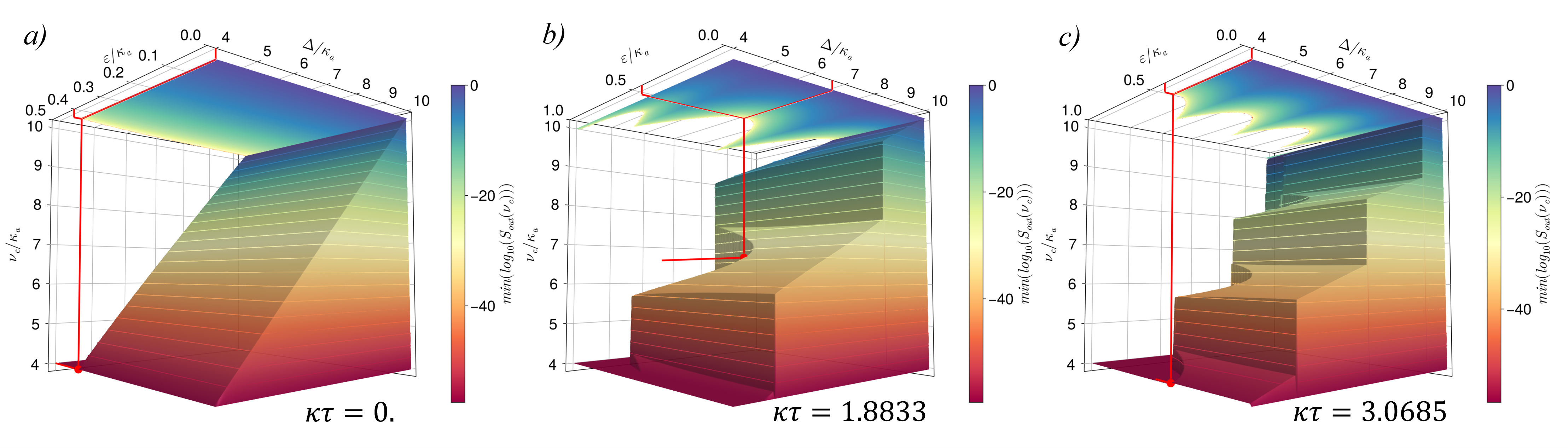}
	\vglue -.2 cm
	\caption{\linespread{1}The best available entanglement at varying mode separation and pump intensity projected to a 2D surface at the top of the figure. The height represents the corresponding frequency as a function of the mode separation and pump amplitude.\lbCap
		Parameters: $\thetap=\pi,\epsilon=0.5\kappa,\kappa_a=\kappa_b=10\cdot2\pi$ MHz, $\kappa_{1,a}=\kappa_{1,b}=0.933\kappa_a, \phi_a=\phi_b=\pi, L_a=L_b=0\%$. }\label{fig:det_eps_one-sided}
	\vglue -.3 cm
\end{figure}

\begin{figure}[h!]
	\centering
	\vglue -.2 cm
	\includegraphics[width=.95\textwidth]{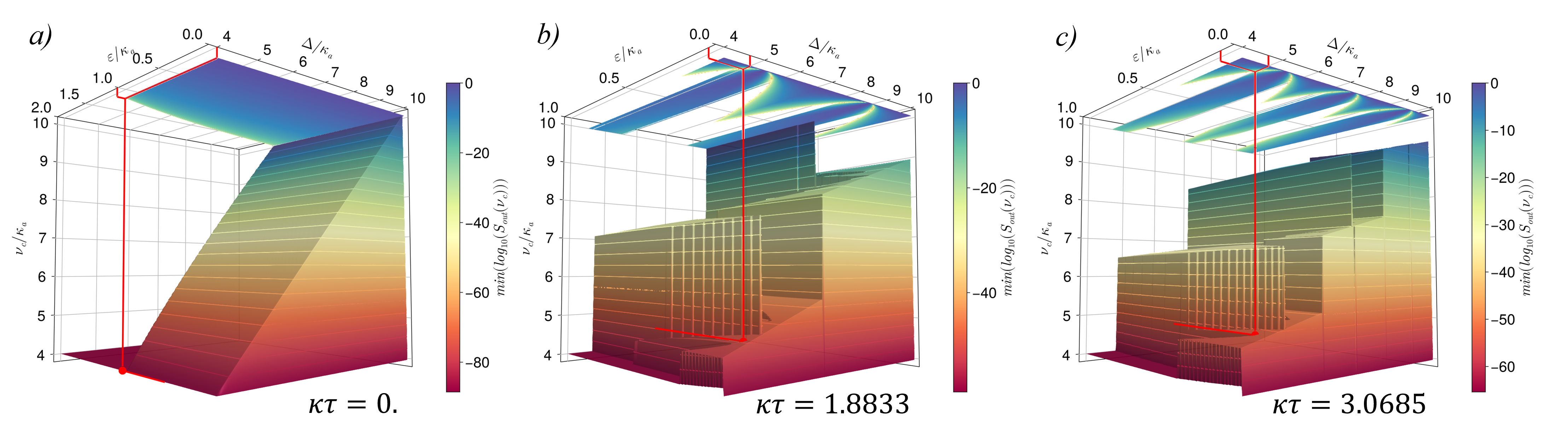}
	\vglue -.2 cm
	\caption{\linespread{1}The best available entanglement at varying mode separation and pump intensity projected to a 2D surface at the top of the figure. The height represents the corresponding frequency as a function of the mode separation and pump amplitude.\lbCap
		Parameters: $\thetap=\pi,\epsilon=0.75\kappa,\Delta=0,\kappa_a=\kappa_b=10\cdot2\pi$ MHz, $\kappa_{1,a}=\kappa_{1,b}=\kappa_a/2,\phi_a=\phi_b=0,L_a=L_b=0\%$. }\label{fig:det_eps_symmetric}
	\vglue -.3 cm
\end{figure}

Considering non-degenerate modes, when all parameters match but the mode frequencies, a periodicity and structure appear in the characteristic frequency as a result of a finite time delay in the feedback loop. The term characteristic is given to this quantity as it characterizes the down-converted field dynamics at the point of the phase transition to the amplification regime. This can also be observed in \reffig{fig:det_eps_one-sided} and \reffig{fig:det_eps_symmetric} as the model becomes invalid where the white spaces are reached in the two-mode squeezing map at the top of the figure. However, previous numerical results showed that the semiclassical calculations follow well these characteristic boundaries.

A key difference between the cases when the crystal is surrounded either by an almost one-sided or by a symmetric resonator is that in the first case a finite pump intensity is required to reach the amplification regime. This is regardless of the length of the time delay in the feedback loop. In the symmetric case, on the other hand, certain combinations of mode separation and time delay enables regimes where the smallest pump intensity can cause immediate amplification and persistent oscillations. 

\clearpage
\section*{References}
\bibliographystyle{apsrev4-2}
\bibliography{NDPA}

\begin{thebibliography}{65}%
\makeatletter
\providecommand \@ifxundefined [1]{%
 \@ifx{#1\undefined}
}%
\providecommand \@ifnum [1]{%
 \ifnum #1\expandafter \@firstoftwo
 \else \expandafter \@secondoftwo
 \fi
}%
\providecommand \@ifx [1]{%
 \ifx #1\expandafter \@firstoftwo
 \else \expandafter \@secondoftwo
 \fi
}%
\providecommand \natexlab [1]{#1}%
\providecommand \enquote  [1]{``#1''}%
\providecommand \bibnamefont  [1]{#1}%
\providecommand \bibfnamefont [1]{#1}%
\providecommand \citenamefont [1]{#1}%
\providecommand \href@noop [0]{\@secondoftwo}%
\providecommand \href [0]{\begingroup \@sanitize@url \@href}%
\providecommand \@href[1]{\@@startlink{#1}\@@href}%
\providecommand \@@href[1]{\endgroup#1\@@endlink}%
\providecommand \@sanitize@url [0]{\catcode `\\12\catcode `\$12\catcode
  `\&12\catcode `\#12\catcode `\^12\catcode `\_12\catcode `\%12\relax}%
\providecommand \@@startlink[1]{}%
\providecommand \@@endlink[0]{}%
\providecommand \url  [0]{\begingroup\@sanitize@url \@url }%
\providecommand \@url [1]{\endgroup\@href {#1}{\urlprefix }}%
\providecommand \urlprefix  [0]{URL }%
\providecommand \Eprint [0]{\href }%
\providecommand \doibase [0]{https://doi.org/}%
\providecommand \selectlanguage [0]{\@gobble}%
\providecommand \bibinfo  [0]{\@secondoftwo}%
\providecommand \bibfield  [0]{\@secondoftwo}%
\providecommand \translation [1]{[#1]}%
\providecommand \BibitemOpen [0]{}%
\providecommand \bibitemStop [0]{}%
\providecommand \bibitemNoStop [0]{.\EOS\space}%
\providecommand \EOS [0]{\spacefactor3000\relax}%
\providecommand \BibitemShut  [1]{\csname bibitem#1\endcsname}%
\let\auto@bib@innerbib\@empty
\bibitem [{\citenamefont {Horodecki}\ \emph {et~al.}(2009)\citenamefont
  {Horodecki}, \citenamefont {Horodecki}, \citenamefont {Horodecki},\ and\
  \citenamefont {Horodecki}}]{Horodecki2007}%
  \BibitemOpen
  \bibfield  {author} {\bibinfo {author} {\bibfnamefont {R.}~\bibnamefont
  {Horodecki}}, \bibinfo {author} {\bibfnamefont {P.}~\bibnamefont
  {Horodecki}}, \bibinfo {author} {\bibfnamefont {M.}~\bibnamefont
  {Horodecki}},\ and\ \bibinfo {author} {\bibfnamefont {K.}~\bibnamefont
  {Horodecki}},\ }\href {https://doi.org/10.1103/RevModPhys.81.865} {\bibfield
  {journal} {\bibinfo  {journal} {Rev. Mod. Phys.}\ }\textbf {\bibinfo {volume}
  {81}},\ \bibinfo {pages} {865} (\bibinfo {year} {2009})}\BibitemShut
  {NoStop}%
\bibitem [{\citenamefont {Eisert}\ and\ \citenamefont
  {Osborne}(2006)}]{Eisert2006}%
  \BibitemOpen
  \bibfield  {author} {\bibinfo {author} {\bibfnamefont {J.}~\bibnamefont
  {Eisert}}\ and\ \bibinfo {author} {\bibfnamefont {T.~J.}\ \bibnamefont
  {Osborne}},\ }\href {https://doi.org/10.1103/PhysRevLett.97.150404}
  {\bibfield  {journal} {\bibinfo  {journal} {Phys. Rev. Lett.}\ }\textbf
  {\bibinfo {volume} {97}},\ \bibinfo {pages} {150404} (\bibinfo {year}
  {2006})}\BibitemShut {NoStop}%
\bibitem [{\citenamefont {Nielsen}\ and\ \citenamefont
  {Chuang}(2010)}]{NielsenChuangBook2010}%
  \BibitemOpen
  \bibfield  {author} {\bibinfo {author} {\bibfnamefont {M.~A.}\ \bibnamefont
  {Nielsen}}\ and\ \bibinfo {author} {\bibfnamefont {I.~L.}\ \bibnamefont
  {Chuang}},\ }\href@noop {} {\emph {\bibinfo {title} {Quantum Computation and
  Quantum Information: 10th Anniversary Edition}}}\ (\bibinfo  {publisher}
  {Cambridge University Press},\ \bibinfo {year} {2010})\BibitemShut {NoStop}%
\bibitem [{\citenamefont {Pichler}\ \emph {et~al.}(2017)\citenamefont
  {Pichler}, \citenamefont {Choi}, \citenamefont {Zoller},\ and\ \citenamefont
  {Lukin}}]{Pichler2017}%
  \BibitemOpen
  \bibfield  {author} {\bibinfo {author} {\bibfnamefont {H.}~\bibnamefont
  {Pichler}}, \bibinfo {author} {\bibfnamefont {S.}~\bibnamefont {Choi}},
  \bibinfo {author} {\bibfnamefont {P.}~\bibnamefont {Zoller}},\ and\ \bibinfo
  {author} {\bibfnamefont {M.~D.}\ \bibnamefont {Lukin}},\ }\href
  {https://doi.org/10.1073/pnas.1711003114} {\bibfield  {journal} {\bibinfo
  {journal} {Proceedings of the National Academy of Sciences}\ }\textbf
  {\bibinfo {volume} {114}},\ \bibinfo {pages} {11362} (\bibinfo {year}
  {2017})},\ \Eprint
  {https://arxiv.org/abs/https://www.pnas.org/doi/pdf/10.1073/pnas.1711003114}
  {https://www.pnas.org/doi/pdf/10.1073/pnas.1711003114} \BibitemShut {NoStop}%
\bibitem [{\citenamefont {Degen}\ \emph {et~al.}(2017)\citenamefont {Degen},
  \citenamefont {Reinhard},\ and\ \citenamefont
  {Cappellaro}}]{QsensingRev2017}%
  \BibitemOpen
  \bibfield  {author} {\bibinfo {author} {\bibfnamefont {C.~L.}\ \bibnamefont
  {Degen}}, \bibinfo {author} {\bibfnamefont {F.}~\bibnamefont {Reinhard}},\
  and\ \bibinfo {author} {\bibfnamefont {P.}~\bibnamefont {Cappellaro}},\
  }\href {https://doi.org/10.1103/RevModPhys.89.035002} {\bibfield  {journal}
  {\bibinfo  {journal} {Rev. Mod. Phys.}\ }\textbf {\bibinfo {volume} {89}},\
  \bibinfo {pages} {035002} (\bibinfo {year} {2017})}\BibitemShut {NoStop}%
\bibitem [{\citenamefont {Glauber}(1963)}]{Glauber1963}%
  \BibitemOpen
  \bibfield  {author} {\bibinfo {author} {\bibfnamefont {R.~J.}\ \bibnamefont
  {Glauber}},\ }\href {https://doi.org/10.1103/PhysRev.130.2529} {\bibfield
  {journal} {\bibinfo  {journal} {Phys. Rev.}\ }\textbf {\bibinfo {volume}
  {130}},\ \bibinfo {pages} {2529} (\bibinfo {year} {1963})}\BibitemShut
  {NoStop}%
\bibitem [{\citenamefont {Breuer}\ and\ \citenamefont
  {Petruccione}(2002)}]{Breuer2002TheTO}%
  \BibitemOpen
  \bibfield  {author} {\bibinfo {author} {\bibfnamefont {H.-P.}\ \bibnamefont
  {Breuer}}\ and\ \bibinfo {author} {\bibfnamefont {F.}~\bibnamefont
  {Petruccione}}\ }(\bibinfo {year} {2002})\BibitemShut {NoStop}%
\bibitem [{\citenamefont {Carmichael}\ and\ \citenamefont
  {Scully}(2003)}]{Carmichael2003StatisticalMI}%
  \BibitemOpen
  \bibfield  {author} {\bibinfo {author} {\bibfnamefont {H.}~\bibnamefont
  {Carmichael}}\ and\ \bibinfo {author} {\bibfnamefont {M.~O.}\ \bibnamefont
  {Scully}}\ }(\bibinfo {year} {2003})\BibitemShut {NoStop}%
\bibitem [{\citenamefont {Koswara}\ \emph {et~al.}(2021)\citenamefont
  {Koswara}, \citenamefont {Bhutoria},\ and\ \citenamefont
  {Chakrabarti}}]{Koswara2021}%
  \BibitemOpen
  \bibfield  {author} {\bibinfo {author} {\bibfnamefont {A.}~\bibnamefont
  {Koswara}}, \bibinfo {author} {\bibfnamefont {V.}~\bibnamefont {Bhutoria}},\
  and\ \bibinfo {author} {\bibfnamefont {R.}~\bibnamefont {Chakrabarti}},\
  }\href {https://doi.org/10.1088/1367-2630/ac0479} {\bibfield  {journal}
  {\bibinfo  {journal} {New Journal of Physics}\ }\textbf {\bibinfo {volume}
  {23}},\ \bibinfo {pages} {063046} (\bibinfo {year} {2021})}\BibitemShut
  {NoStop}%
\bibitem [{\citenamefont {Daoyi~Dong}(2023)}]{DongPetersenBook2023}%
  \BibitemOpen
  \bibfield  {author} {\bibinfo {author} {\bibfnamefont {I.~R.~P.}\
  \bibnamefont {Daoyi~Dong}},\ }in\ \href
  {https://doi.org/10.1007/978-3-031-20245-2} {\emph {\bibinfo {booktitle}
  {Communications and Control Engineering}}}\ (\bibinfo {year}
  {2023})\BibitemShut {NoStop}%
\bibitem [{\citenamefont {Zhang}\ \emph {et~al.}(2017)\citenamefont {Zhang},
  \citenamefont {xi~Liu}, \citenamefont {Wu}, \citenamefont {Jacobs},\ and\
  \citenamefont {Nori}}]{Zhang2017}%
  \BibitemOpen
  \bibfield  {author} {\bibinfo {author} {\bibfnamefont {J.}~\bibnamefont
  {Zhang}}, \bibinfo {author} {\bibfnamefont {Y.}~\bibnamefont {xi~Liu}},
  \bibinfo {author} {\bibfnamefont {R.-B.}\ \bibnamefont {Wu}}, \bibinfo
  {author} {\bibfnamefont {K.}~\bibnamefont {Jacobs}},\ and\ \bibinfo {author}
  {\bibfnamefont {F.}~\bibnamefont {Nori}},\ }\href
  {https://doi.org/https://doi.org/10.1016/j.physrep.2017.02.003} {\bibfield
  {journal} {\bibinfo  {journal} {Physics Reports}\ }\textbf {\bibinfo {volume}
  {679}},\ \bibinfo {pages} {1} (\bibinfo {year} {2017})},\ \bibinfo {note}
  {quantum feedback: theory, experiments, and applications}\BibitemShut
  {NoStop}%
\bibitem [{\citenamefont {Wiseman}\ and\ \citenamefont
  {Milburn}(2010)}]{Wiseman2010}%
  \BibitemOpen
  \bibfield  {author} {\bibinfo {author} {\bibfnamefont {H.~M.}\ \bibnamefont
  {Wiseman}}\ and\ \bibinfo {author} {\bibfnamefont {G.~J.}\ \bibnamefont
  {Milburn}},\ }\href {https://doi.org/10.1017/CBO9780511813948} {\emph
  {\bibinfo {title} {Quantum Measurement and Control}}}\ (\bibinfo  {publisher}
  {Cambridge University Press},\ \bibinfo {year} {2010})\BibitemShut {NoStop}%
\bibitem [{\citenamefont {Eschner}\ \emph {et~al.}(2001)\citenamefont
  {Eschner}, \citenamefont {Raab}, \citenamefont {Schmidt-Kaler},\ and\
  \citenamefont {Blatt}}]{Eschner2001}%
  \BibitemOpen
  \bibfield  {author} {\bibinfo {author} {\bibfnamefont {J.}~\bibnamefont
  {Eschner}}, \bibinfo {author} {\bibfnamefont {C.}~\bibnamefont {Raab}},
  \bibinfo {author} {\bibfnamefont {F.}~\bibnamefont {Schmidt-Kaler}},\ and\
  \bibinfo {author} {\bibfnamefont {R.}~\bibnamefont {Blatt}},\ }\href
  {https://doi.org/10.1038/35097017} {\bibfield  {journal} {\bibinfo  {journal}
  {Nature}\ }\textbf {\bibinfo {volume} {413}},\ \bibinfo {pages} {495}
  (\bibinfo {year} {2001})}\BibitemShut {NoStop}%
\bibitem [{\citenamefont {Wilson}\ \emph {et~al.}(2003)\citenamefont {Wilson},
  \citenamefont {Bushev}, \citenamefont {Eschner}, \citenamefont
  {Schmidt-Kaler}, \citenamefont {Becher}, \citenamefont {Blatt},\ and\
  \citenamefont {Dorner}}]{Wilson2003}%
  \BibitemOpen
  \bibfield  {author} {\bibinfo {author} {\bibfnamefont {M.~A.}\ \bibnamefont
  {Wilson}}, \bibinfo {author} {\bibfnamefont {P.}~\bibnamefont {Bushev}},
  \bibinfo {author} {\bibfnamefont {J.}~\bibnamefont {Eschner}}, \bibinfo
  {author} {\bibfnamefont {F.}~\bibnamefont {Schmidt-Kaler}}, \bibinfo {author}
  {\bibfnamefont {C.}~\bibnamefont {Becher}}, \bibinfo {author} {\bibfnamefont
  {R.}~\bibnamefont {Blatt}},\ and\ \bibinfo {author} {\bibfnamefont
  {U.}~\bibnamefont {Dorner}},\ }\href
  {https://doi.org/10.1103/PhysRevLett.91.213602} {\bibfield  {journal}
  {\bibinfo  {journal} {Phys. Rev. Lett.}\ }\textbf {\bibinfo {volume} {91}},\
  \bibinfo {pages} {213602} (\bibinfo {year} {2003})}\BibitemShut {NoStop}%
\bibitem [{\citenamefont {Dubin}\ \emph {et~al.}(2007)\citenamefont {Dubin},
  \citenamefont {Rotter}, \citenamefont {Mukherjee}, \citenamefont {Russo},
  \citenamefont {Eschner},\ and\ \citenamefont {Blatt}}]{Dubin2007}%
  \BibitemOpen
  \bibfield  {author} {\bibinfo {author} {\bibfnamefont {F.~m.~c.}\
  \bibnamefont {Dubin}}, \bibinfo {author} {\bibfnamefont {D.}~\bibnamefont
  {Rotter}}, \bibinfo {author} {\bibfnamefont {M.}~\bibnamefont {Mukherjee}},
  \bibinfo {author} {\bibfnamefont {C.}~\bibnamefont {Russo}}, \bibinfo
  {author} {\bibfnamefont {J.}~\bibnamefont {Eschner}},\ and\ \bibinfo {author}
  {\bibfnamefont {R.}~\bibnamefont {Blatt}},\ }\href
  {https://doi.org/10.1103/PhysRevLett.98.183003} {\bibfield  {journal}
  {\bibinfo  {journal} {Phys. Rev. Lett.}\ }\textbf {\bibinfo {volume} {98}},\
  \bibinfo {pages} {183003} (\bibinfo {year} {2007})}\BibitemShut {NoStop}%
\bibitem [{\citenamefont {Mabuchi}(2008)}]{Mabuchi2008}%
  \BibitemOpen
  \bibfield  {author} {\bibinfo {author} {\bibfnamefont {H.}~\bibnamefont
  {Mabuchi}},\ }\href {https://doi.org/10.1103/PhysRevA.78.032323} {\bibfield
  {journal} {\bibinfo  {journal} {Phys. Rev. A - At. Mol. Opt. Phys.}\ }\textbf
  {\bibinfo {volume} {78}},\ \bibinfo {pages} {032323} (\bibinfo {year}
  {2008})},\ \Eprint {https://arxiv.org/abs/0803.2007} {arXiv:0803.2007}
  \BibitemShut {NoStop}%
\bibitem [{\citenamefont {Iida}\ \emph {et~al.}(2012)\citenamefont {Iida},
  \citenamefont {Yukawa}, \citenamefont {Yonezawa}, \citenamefont {Yamamoto},\
  and\ \citenamefont {Furusawa}}]{Iida2012}%
  \BibitemOpen
  \bibfield  {author} {\bibinfo {author} {\bibfnamefont {S.}~\bibnamefont
  {Iida}}, \bibinfo {author} {\bibfnamefont {M.}~\bibnamefont {Yukawa}},
  \bibinfo {author} {\bibfnamefont {H.}~\bibnamefont {Yonezawa}}, \bibinfo
  {author} {\bibfnamefont {N.}~\bibnamefont {Yamamoto}},\ and\ \bibinfo
  {author} {\bibfnamefont {A.}~\bibnamefont {Furusawa}},\ }\href
  {https://doi.org/10.1109/TAC.2012.2195831} {\bibfield  {journal} {\bibinfo
  {journal} {IEEE Trans. Automat. Contr.}\ }\textbf {\bibinfo {volume} {57}},\
  \bibinfo {pages} {2045} (\bibinfo {year} {2012})},\ \Eprint
  {https://arxiv.org/abs/1103.1324} {arXiv:1103.1324} \BibitemShut {NoStop}%
\bibitem [{\citenamefont {Jacobs}\ \emph {et~al.}(2014)\citenamefont {Jacobs},
  \citenamefont {Wang},\ and\ \citenamefont {Wiseman}}]{Jacobs2014}%
  \BibitemOpen
  \bibfield  {author} {\bibinfo {author} {\bibfnamefont {K.}~\bibnamefont
  {Jacobs}}, \bibinfo {author} {\bibfnamefont {X.}~\bibnamefont {Wang}},\ and\
  \bibinfo {author} {\bibfnamefont {H.~M.}\ \bibnamefont {Wiseman}},\ }\href
  {https://doi.org/10.1088/1367-2630/16/7/073036} {\bibfield  {journal}
  {\bibinfo  {journal} {New Journal of Physics}\ }\textbf {\bibinfo {volume}
  {16}},\ \bibinfo {pages} {073036} (\bibinfo {year} {2014})}\BibitemShut
  {NoStop}%
\bibitem [{\citenamefont {Dong}\ and\ \citenamefont
  {Petersen}(2010)}]{Dong2010}%
  \BibitemOpen
  \bibfield  {author} {\bibinfo {author} {\bibfnamefont {D.}~\bibnamefont
  {Dong}}\ and\ \bibinfo {author} {\bibfnamefont {I.}~\bibnamefont
  {Petersen}},\ }\href
  {https://digital-library.theiet.org/content/journals/10.1049/iet-cta.2009.0508}
  {\bibfield  {journal} {\bibinfo  {journal} {IET Control Theory and
  Applications}\ }\textbf {\bibinfo {volume} {4}},\ \bibinfo {pages} {2651}
  (\bibinfo {year} {2010})}\BibitemShut {NoStop}%
\bibitem [{\citenamefont {Geffert}(2015)}]{Geffert2015}%
  \BibitemOpen
  \bibfield  {author} {\bibinfo {author} {\bibfnamefont {P.~M.}\ \bibnamefont
  {Geffert}},\ }\href@noop {} {\emph {\bibinfo {title} {Stochastic
  Non-Excitable Systems with Time Delay: Modulation of Noise Effects by
  Time-Delayed Feedback}}},\ BestMasters\ (\bibinfo  {publisher} {Springer},\
  \bibinfo {address} {Wiesbaden},\ \bibinfo {year} {2015})\BibitemShut
  {NoStop}%
\bibitem [{\citenamefont {Jaurigue}(2017)}]{Jaurigue2017}%
  \BibitemOpen
  \bibfield  {author} {\bibinfo {author} {\bibfnamefont {L.}~\bibnamefont
  {Jaurigue}},\ }\href@noop {} {\emph {\bibinfo {title} {Passively Mode-Locked
  Semiconductor Lasers: Dynamics and Stochastic Properties in the Presence of
  Optical Feedback}}},\ Springer Thesis\ (\bibinfo  {publisher} {Springer},\
  \bibinfo {address} {Cham},\ \bibinfo {year} {2017})\BibitemShut {NoStop}%
\bibitem [{\citenamefont {Kabuss}\ \emph {et~al.}(2016)\citenamefont {Kabuss},
  \citenamefont {Katsch}, \citenamefont {Knorr},\ and\ \citenamefont
  {Carmele}}]{Kabuss2016}%
  \BibitemOpen
  \bibfield  {author} {\bibinfo {author} {\bibfnamefont {J.}~\bibnamefont
  {Kabuss}}, \bibinfo {author} {\bibfnamefont {F.}~\bibnamefont {Katsch}},
  \bibinfo {author} {\bibfnamefont {A.}~\bibnamefont {Knorr}},\ and\ \bibinfo
  {author} {\bibfnamefont {A.}~\bibnamefont {Carmele}},\ }\href
  {https://doi.org/10.1364/JOSAB.33.000C10} {\bibfield  {journal} {\bibinfo
  {journal} {J. Opt. Soc. Am. B}\ }\textbf {\bibinfo {volume} {33}},\ \bibinfo
  {pages} {C10} (\bibinfo {year} {2016})}\BibitemShut {NoStop}%
\bibitem [{\citenamefont {{Pyragas K.}}(1992)}]{Pyragas1992}%
  \BibitemOpen
  \bibfield  {author} {\bibinfo {author} {\bibnamefont {{Pyragas K.}}},\ }\href
  {https://doi.org/10.1016/0375-9601(92)90745-8} {\bibfield  {journal}
  {\bibinfo  {journal} {Phys. Lett. A}\ }\textbf {\bibinfo {volume} {170}},\
  \bibinfo {pages} {421} (\bibinfo {year} {1992})}\BibitemShut {NoStop}%
\bibitem [{\citenamefont {Droenner}\ \emph {et~al.}(2019)\citenamefont
  {Droenner}, \citenamefont {Naumann}, \citenamefont {Sch\"oll}, \citenamefont
  {Knorr},\ and\ \citenamefont {Carmele}}]{Droenner2019}%
  \BibitemOpen
  \bibfield  {author} {\bibinfo {author} {\bibfnamefont {L.}~\bibnamefont
  {Droenner}}, \bibinfo {author} {\bibfnamefont {N.~L.}\ \bibnamefont
  {Naumann}}, \bibinfo {author} {\bibfnamefont {E.}~\bibnamefont {Sch\"oll}},
  \bibinfo {author} {\bibfnamefont {A.}~\bibnamefont {Knorr}},\ and\ \bibinfo
  {author} {\bibfnamefont {A.}~\bibnamefont {Carmele}},\ }\href
  {https://doi.org/10.1103/PhysRevA.99.023840} {\bibfield  {journal} {\bibinfo
  {journal} {Phys. Rev. A}\ }\textbf {\bibinfo {volume} {99}},\ \bibinfo
  {pages} {023840} (\bibinfo {year} {2019})}\BibitemShut {NoStop}%
\bibitem [{\citenamefont {Naumann}\ \emph {et~al.}(2014)\citenamefont
  {Naumann}, \citenamefont {Hein}, \citenamefont {Knorr},\ and\ \citenamefont
  {Kabuss}}]{Naumann2014}%
  \BibitemOpen
  \bibfield  {author} {\bibinfo {author} {\bibfnamefont {N.~L.}\ \bibnamefont
  {Naumann}}, \bibinfo {author} {\bibfnamefont {S.~M.}\ \bibnamefont {Hein}},
  \bibinfo {author} {\bibfnamefont {A.}~\bibnamefont {Knorr}},\ and\ \bibinfo
  {author} {\bibfnamefont {J.}~\bibnamefont {Kabuss}},\ }\href
  {https://doi.org/10.1103/PhysRevA.90.043835} {\bibfield  {journal} {\bibinfo
  {journal} {Phys. Rev. A}\ }\textbf {\bibinfo {volume} {90}},\ \bibinfo
  {pages} {043835} (\bibinfo {year} {2014})}\BibitemShut {NoStop}%
\bibitem [{\citenamefont {Grimsmo}\ \emph {et~al.}(2014)\citenamefont
  {Grimsmo}, \citenamefont {Parkins},\ and\ \citenamefont
  {Skagerstam}}]{Grimsmo2014}%
  \BibitemOpen
  \bibfield  {author} {\bibinfo {author} {\bibfnamefont {A.~L.}\ \bibnamefont
  {Grimsmo}}, \bibinfo {author} {\bibfnamefont {A.~S.}\ \bibnamefont
  {Parkins}},\ and\ \bibinfo {author} {\bibfnamefont {B.~S.}\ \bibnamefont
  {Skagerstam}},\ }\href {https://doi.org/10.1088/1367-2630/16/6/065004}
  {\bibfield  {journal} {\bibinfo  {journal} {New J. Phys.}\ }\textbf {\bibinfo
  {volume} {16}},\ \bibinfo {pages} {065004} (\bibinfo {year} {2014})},\
  \Eprint {https://arxiv.org/abs/1401.2287} {arXiv:1401.2287} \BibitemShut
  {NoStop}%
\bibitem [{\citenamefont {Carmele}\ \emph {et~al.}(2013)\citenamefont
  {Carmele}, \citenamefont {Kabuss}, \citenamefont {Schulze}, \citenamefont
  {Reitzenstein},\ and\ \citenamefont {Knorr}}]{Carmele2013}%
  \BibitemOpen
  \bibfield  {author} {\bibinfo {author} {\bibfnamefont {A.}~\bibnamefont
  {Carmele}}, \bibinfo {author} {\bibfnamefont {J.}~\bibnamefont {Kabuss}},
  \bibinfo {author} {\bibfnamefont {F.}~\bibnamefont {Schulze}}, \bibinfo
  {author} {\bibfnamefont {S.}~\bibnamefont {Reitzenstein}},\ and\ \bibinfo
  {author} {\bibfnamefont {A.}~\bibnamefont {Knorr}},\ }\href
  {https://doi.org/10.1103/PhysRevLett.110.013601} {\bibfield  {journal}
  {\bibinfo  {journal} {Phys. Rev. Lett.}\ }\textbf {\bibinfo {volume} {110}},\
  \bibinfo {pages} {013601} (\bibinfo {year} {2013})}\BibitemShut {NoStop}%
\bibitem [{\citenamefont {Kabuss}\ \emph {et~al.}(2015)\citenamefont {Kabuss},
  \citenamefont {Krimer}, \citenamefont {Rotter}, \citenamefont {Stannigel},
  \citenamefont {Knorr},\ and\ \citenamefont {Carmele}}]{Kabuss2015}%
  \BibitemOpen
  \bibfield  {author} {\bibinfo {author} {\bibfnamefont {J.}~\bibnamefont
  {Kabuss}}, \bibinfo {author} {\bibfnamefont {D.~O.}\ \bibnamefont {Krimer}},
  \bibinfo {author} {\bibfnamefont {S.}~\bibnamefont {Rotter}}, \bibinfo
  {author} {\bibfnamefont {K.}~\bibnamefont {Stannigel}}, \bibinfo {author}
  {\bibfnamefont {A.}~\bibnamefont {Knorr}},\ and\ \bibinfo {author}
  {\bibfnamefont {A.}~\bibnamefont {Carmele}},\ }\href
  {https://doi.org/10.1103/PhysRevA.92.053801} {\bibfield  {journal} {\bibinfo
  {journal} {Phys. Rev. A}\ }\textbf {\bibinfo {volume} {92}},\ \bibinfo
  {pages} {053801} (\bibinfo {year} {2015})}\BibitemShut {NoStop}%
\bibitem [{\citenamefont {Kraft}\ \emph {et~al.}(2016)\citenamefont {Kraft},
  \citenamefont {Hein}, \citenamefont {Lehnert}, \citenamefont {Sch{\"{o}}ll},
  \citenamefont {Hughes},\ and\ \citenamefont {Knorr}}]{Kraft2016}%
  \BibitemOpen
  \bibfield  {author} {\bibinfo {author} {\bibfnamefont {M.}~\bibnamefont
  {Kraft}}, \bibinfo {author} {\bibfnamefont {S.~M.}\ \bibnamefont {Hein}},
  \bibinfo {author} {\bibfnamefont {J.}~\bibnamefont {Lehnert}}, \bibinfo
  {author} {\bibfnamefont {E.}~\bibnamefont {Sch{\"{o}}ll}}, \bibinfo {author}
  {\bibfnamefont {S.}~\bibnamefont {Hughes}},\ and\ \bibinfo {author}
  {\bibfnamefont {A.}~\bibnamefont {Knorr}},\ }\href
  {https://doi.org/10.1103/PhysRevA.94.023806} {\bibfield  {journal} {\bibinfo
  {journal} {Phys. Rev. A}\ }\textbf {\bibinfo {volume} {94}},\ \bibinfo
  {pages} {023806} (\bibinfo {year} {2016})},\ \Eprint
  {https://arxiv.org/abs/1603.07137v1} {arXiv:1603.07137v1} \BibitemShut
  {NoStop}%
\bibitem [{\citenamefont {N{\'{e}}met}\ and\ \citenamefont
  {Parkins}(2016)}]{Nemet2016}%
  \BibitemOpen
  \bibfield  {author} {\bibinfo {author} {\bibfnamefont {N.}~\bibnamefont
  {N{\'{e}}met}}\ and\ \bibinfo {author} {\bibfnamefont {S.}~\bibnamefont
  {Parkins}},\ }\href {https://doi.org/10.1103/PhysRevA.94.023809} {\bibfield
  {journal} {\bibinfo  {journal} {Phys. Rev. A}\ }\textbf {\bibinfo {volume}
  {94}},\ \bibinfo {pages} {023809} (\bibinfo {year} {2016})},\ \Eprint
  {https://arxiv.org/abs/1606.00178} {arXiv:1606.00178} \BibitemShut {NoStop}%
\bibitem [{\citenamefont {Gough}\ and\ \citenamefont
  {Wildfeuer}(2009)}]{Gough2009}%
  \BibitemOpen
  \bibfield  {author} {\bibinfo {author} {\bibfnamefont {J.~E.}\ \bibnamefont
  {Gough}}\ and\ \bibinfo {author} {\bibfnamefont {S.}~\bibnamefont
  {Wildfeuer}},\ }\href {https://doi.org/10.1103/PhysRevA.80.042107} {\bibfield
   {journal} {\bibinfo  {journal} {Phys. Rev. A}\ }\textbf {\bibinfo {volume}
  {80}},\ \bibinfo {pages} {042107} (\bibinfo {year} {2009})}\BibitemShut
  {NoStop}%
\bibitem [{\citenamefont {Crisafulli}\ \emph {et~al.}(2013)\citenamefont
  {Crisafulli}, \citenamefont {Tezak}, \citenamefont {Soh}, \citenamefont
  {Armen},\ and\ \citenamefont {Mabuchi}}]{Crisafulli2013}%
  \BibitemOpen
  \bibfield  {author} {\bibinfo {author} {\bibfnamefont {O.}~\bibnamefont
  {Crisafulli}}, \bibinfo {author} {\bibfnamefont {N.}~\bibnamefont {Tezak}},
  \bibinfo {author} {\bibfnamefont {D.~B.~S.}\ \bibnamefont {Soh}}, \bibinfo
  {author} {\bibfnamefont {M.~a.}\ \bibnamefont {Armen}},\ and\ \bibinfo
  {author} {\bibfnamefont {H.}~\bibnamefont {Mabuchi}},\ }\href
  {https://doi.org/10.1364/OE.21.018371} {\bibfield  {journal} {\bibinfo
  {journal} {Opt. Express}\ }\textbf {\bibinfo {volume} {21}},\ \bibinfo
  {pages} {18371} (\bibinfo {year} {2013})},\ \Eprint
  {https://arxiv.org/abs/1302.6179} {arXiv:1302.6179} \BibitemShut {NoStop}%
\bibitem [{\citenamefont {Hein}\ \emph {et~al.}(2014)\citenamefont {Hein},
  \citenamefont {Schulze}, \citenamefont {Carmele},\ and\ \citenamefont
  {Knorr}}]{Hein2014}%
  \BibitemOpen
  \bibfield  {author} {\bibinfo {author} {\bibfnamefont {S.~M.}\ \bibnamefont
  {Hein}}, \bibinfo {author} {\bibfnamefont {F.}~\bibnamefont {Schulze}},
  \bibinfo {author} {\bibfnamefont {A.}~\bibnamefont {Carmele}},\ and\ \bibinfo
  {author} {\bibfnamefont {A.}~\bibnamefont {Knorr}},\ }\href
  {https://doi.org/10.1103/PhysRevLett.113.027401} {\bibfield  {journal}
  {\bibinfo  {journal} {Phys. Rev. Lett.}\ }\textbf {\bibinfo {volume} {113}},\
  \bibinfo {pages} {027401} (\bibinfo {year} {2014})}\BibitemShut {NoStop}%
\bibitem [{\citenamefont {Hein}\ \emph {et~al.}(2015)\citenamefont {Hein},
  \citenamefont {Schulze}, \citenamefont {Carmele},\ and\ \citenamefont
  {Knorr}}]{Hein2015}%
  \BibitemOpen
  \bibfield  {author} {\bibinfo {author} {\bibfnamefont {S.~M.}\ \bibnamefont
  {Hein}}, \bibinfo {author} {\bibfnamefont {F.}~\bibnamefont {Schulze}},
  \bibinfo {author} {\bibfnamefont {A.}~\bibnamefont {Carmele}},\ and\ \bibinfo
  {author} {\bibfnamefont {A.}~\bibnamefont {Knorr}},\ }\href
  {https://doi.org/10.1103/PhysRevA.91.052321} {\bibfield  {journal} {\bibinfo
  {journal} {Phys. Rev. A}\ }\textbf {\bibinfo {volume} {91}},\ \bibinfo
  {pages} {052321} (\bibinfo {year} {2015})}\BibitemShut {NoStop}%
\bibitem [{\citenamefont {Hein}\ \emph {et~al.}(2016)\citenamefont {Hein},
  \citenamefont {Carmele},\ and\ \citenamefont {Knorr}}]{Hein2016a}%
  \BibitemOpen
  \bibfield  {author} {\bibinfo {author} {\bibfnamefont {S.~M.}\ \bibnamefont
  {Hein}}, \bibinfo {author} {\bibfnamefont {A.}~\bibnamefont {Carmele}},\ and\
  \bibinfo {author} {\bibfnamefont {A.}~\bibnamefont {Knorr}},\ }in\ \href
  {https://doi.org/10.1117/12.2207671} {\emph {\bibinfo {booktitle} {Proc. SPIE
  9742, Phys. Simul. Optoelectron. Devices XXIV}}},\ \bibinfo {series and
  number} {\bibinfo {number} {March 2016}}\ (\bibinfo {year} {2016})\ p.\
  \bibinfo {pages} {97420X}\BibitemShut {NoStop}%
\bibitem [{\citenamefont {N\'emet}\ \emph {et~al.}(2019)\citenamefont
  {N\'emet}, \citenamefont {Parkins}, \citenamefont {Knorr},\ and\
  \citenamefont {Carmele}}]{Nemet2019}%
  \BibitemOpen
  \bibfield  {author} {\bibinfo {author} {\bibfnamefont {N.}~\bibnamefont
  {N\'emet}}, \bibinfo {author} {\bibfnamefont {S.}~\bibnamefont {Parkins}},
  \bibinfo {author} {\bibfnamefont {A.}~\bibnamefont {Knorr}},\ and\ \bibinfo
  {author} {\bibfnamefont {A.}~\bibnamefont {Carmele}},\ }\href
  {https://doi.org/10.1103/PhysRevA.99.053809} {\bibfield  {journal} {\bibinfo
  {journal} {Phys. Rev. A}\ }\textbf {\bibinfo {volume} {99}},\ \bibinfo
  {pages} {053809} (\bibinfo {year} {2019})}\BibitemShut {NoStop}%
\bibitem [{\citenamefont {Schmid}\ \emph {et~al.}(2022)\citenamefont {Schmid},
  \citenamefont {Ngai}, \citenamefont {Ernzer}, \citenamefont {Aguilera},
  \citenamefont {Karg},\ and\ \citenamefont {Treutlein}}]{Schmid2022}%
  \BibitemOpen
  \bibfield  {author} {\bibinfo {author} {\bibfnamefont {G.-L.}\ \bibnamefont
  {Schmid}}, \bibinfo {author} {\bibfnamefont {C.~T.}\ \bibnamefont {Ngai}},
  \bibinfo {author} {\bibfnamefont {M.}~\bibnamefont {Ernzer}}, \bibinfo
  {author} {\bibfnamefont {M.~B.}\ \bibnamefont {Aguilera}}, \bibinfo {author}
  {\bibfnamefont {T.~M.}\ \bibnamefont {Karg}},\ and\ \bibinfo {author}
  {\bibfnamefont {P.}~\bibnamefont {Treutlein}},\ }\href
  {https://doi.org/10.1103/PhysRevX.12.011020} {\bibfield  {journal} {\bibinfo
  {journal} {Phys. Rev. X}\ }\textbf {\bibinfo {volume} {12}},\ \bibinfo
  {pages} {011020} (\bibinfo {year} {2022})}\BibitemShut {NoStop}%
\bibitem [{\citenamefont {Ernzer}\ \emph {et~al.}(2023)\citenamefont {Ernzer},
  \citenamefont {Bosch~Aguilera}, \citenamefont {Brunelli}, \citenamefont
  {Schmid}, \citenamefont {Karg}, \citenamefont {Bruder}, \citenamefont
  {Potts},\ and\ \citenamefont {Treutlein}}]{Ernzer2023}%
  \BibitemOpen
  \bibfield  {author} {\bibinfo {author} {\bibfnamefont {M.}~\bibnamefont
  {Ernzer}}, \bibinfo {author} {\bibfnamefont {M.}~\bibnamefont
  {Bosch~Aguilera}}, \bibinfo {author} {\bibfnamefont {M.}~\bibnamefont
  {Brunelli}}, \bibinfo {author} {\bibfnamefont {G.-L.}\ \bibnamefont
  {Schmid}}, \bibinfo {author} {\bibfnamefont {T.~M.}\ \bibnamefont {Karg}},
  \bibinfo {author} {\bibfnamefont {C.}~\bibnamefont {Bruder}}, \bibinfo
  {author} {\bibfnamefont {P.~P.}\ \bibnamefont {Potts}},\ and\ \bibinfo
  {author} {\bibfnamefont {P.}~\bibnamefont {Treutlein}},\ }\href
  {https://doi.org/10.1103/PhysRevX.13.021023} {\bibfield  {journal} {\bibinfo
  {journal} {Phys. Rev. X}\ }\textbf {\bibinfo {volume} {13}},\ \bibinfo
  {pages} {021023} (\bibinfo {year} {2023})}\BibitemShut {NoStop}%
\bibitem [{\citenamefont {Collett}\ and\ \citenamefont
  {Loudon}(1987)}]{Collett1987}%
  \BibitemOpen
  \bibfield  {author} {\bibinfo {author} {\bibfnamefont {M.~J.}\ \bibnamefont
  {Collett}}\ and\ \bibinfo {author} {\bibfnamefont {R.}~\bibnamefont
  {Loudon}},\ }\href {https://doi.org/10.1364/JOSAB.4.001525} {\bibfield
  {journal} {\bibinfo  {journal} {J. Opt. Soc. Am. B}\ }\textbf {\bibinfo
  {volume} {4}},\ \bibinfo {pages} {1525} (\bibinfo {year} {1987})}\BibitemShut
  {NoStop}%
\bibitem [{\citenamefont {Drummond}\ and\ \citenamefont
  {Reid}(1990)}]{Drummond1990}%
  \BibitemOpen
  \bibfield  {author} {\bibinfo {author} {\bibfnamefont {P.~D.}\ \bibnamefont
  {Drummond}}\ and\ \bibinfo {author} {\bibfnamefont {M.~D.}\ \bibnamefont
  {Reid}},\ }\href {https://doi.org/10.1103/PhysRevA.41.3930} {\bibfield
  {journal} {\bibinfo  {journal} {Phys. Rev. A}\ }\textbf {\bibinfo {volume}
  {41}},\ \bibinfo {pages} {3930} (\bibinfo {year} {1990})}\BibitemShut
  {NoStop}%
\bibitem [{\citenamefont {Ou}\ \emph {et~al.}(1992)\citenamefont {Ou},
  \citenamefont {Pereira},\ and\ \citenamefont {Kimble}}]{Ou1992}%
  \BibitemOpen
  \bibfield  {author} {\bibinfo {author} {\bibfnamefont {Z.~Y.}\ \bibnamefont
  {Ou}}, \bibinfo {author} {\bibfnamefont {S.~F.}\ \bibnamefont {Pereira}},\
  and\ \bibinfo {author} {\bibfnamefont {H.~J.}\ \bibnamefont {Kimble}},\
  }\href {https://doi.org/10.1007/BF00325015} {\bibfield  {journal} {\bibinfo
  {journal} {Appl. Phys. B}\ }\textbf {\bibinfo {volume} {55}},\ \bibinfo
  {pages} {265} (\bibinfo {year} {1992})}\BibitemShut {NoStop}%
\bibitem [{\citenamefont {Braunstein}\ and\ \citenamefont
  {Pati}(2003)}]{Braunstein2003}%
  \BibitemOpen
  \bibfield  {author} {\bibinfo {author} {\bibfnamefont {S.~L.}\ \bibnamefont
  {Braunstein}}\ and\ \bibinfo {author} {\bibfnamefont {A.~K.}\ \bibnamefont
  {Pati}},\ }\href {https://doi.org/10.1007/978-94-015-1258-9} {\emph {\bibinfo
  {title} {{Quantum Information with Continuous Variables}}}}\ (\bibinfo
  {publisher} {Springer},\ \bibinfo {year} {2003})\BibitemShut {NoStop}%
\bibitem [{\citenamefont {Braunstein}\ and\ \citenamefont {van
  Loock}(2005)}]{Braunstein2005}%
  \BibitemOpen
  \bibfield  {author} {\bibinfo {author} {\bibfnamefont {S.~L.}\ \bibnamefont
  {Braunstein}}\ and\ \bibinfo {author} {\bibfnamefont {P.}~\bibnamefont {van
  Loock}},\ }\href {https://doi.org/10.1103/RevModPhys.77.513} {\bibfield
  {journal} {\bibinfo  {journal} {Rev. Mod. Phys.}\ }\textbf {\bibinfo {volume}
  {77}},\ \bibinfo {pages} {513} (\bibinfo {year} {2005})}\BibitemShut
  {NoStop}%
\bibitem [{\citenamefont {He}\ and\ \citenamefont {Li}(2007)}]{He2007}%
  \BibitemOpen
  \bibfield  {author} {\bibinfo {author} {\bibfnamefont {W.-P.~P.}\
  \bibnamefont {He}}\ and\ \bibinfo {author} {\bibfnamefont {F.-L.~L.}\
  \bibnamefont {Li}},\ }\href {https://doi.org/10.1103/PhysRevA.76.012328}
  {\bibfield  {journal} {\bibinfo  {journal} {Phys. Rev. A}\ }\textbf {\bibinfo
  {volume} {76}},\ \bibinfo {pages} {012328} (\bibinfo {year}
  {2007})}\BibitemShut {NoStop}%
\bibitem [{\citenamefont {Shi}\ and\ \citenamefont {Nurdin}(2016)}]{Shi2016a}%
  \BibitemOpen
  \bibfield  {author} {\bibinfo {author} {\bibfnamefont {Z.}~\bibnamefont
  {Shi}}\ and\ \bibinfo {author} {\bibfnamefont {H.~I.}\ \bibnamefont
  {Nurdin}},\ }\href {https://doi.org/10.1109/ACC.2016.7526108} {\bibfield
  {journal} {\bibinfo  {journal} {Proc. Am. Control Conf.}\ }\textbf {\bibinfo
  {volume} {2016-July}},\ \bibinfo {pages} {1} (\bibinfo {year} {2016})},\
  \Eprint {https://arxiv.org/abs/arXiv:1603.03114v2} {arXiv:arXiv:1603.03114v2}
  \BibitemShut {NoStop}%
\bibitem [{\citenamefont {Li}\ \emph {et~al.}(2006)\citenamefont {Li},
  \citenamefont {Tan},\ and\ \citenamefont {Ke}}]{Li2006}%
  \BibitemOpen
  \bibfield  {author} {\bibinfo {author} {\bibfnamefont {G.~X.}\ \bibnamefont
  {Li}}, \bibinfo {author} {\bibfnamefont {H.~T.}\ \bibnamefont {Tan}},\ and\
  \bibinfo {author} {\bibfnamefont {S.~S.}\ \bibnamefont {Ke}},\ }\href
  {https://doi.org/10.1103/PhysRevA.74.012304} {\bibfield  {journal} {\bibinfo
  {journal} {Phys. Rev. A}\ }\textbf {\bibinfo {volume} {74}},\ \bibinfo
  {pages} {012304} (\bibinfo {year} {2006})}\BibitemShut {NoStop}%
\bibitem [{\citenamefont {Ke}\ \emph {et~al.}(2007)\citenamefont {Ke},
  \citenamefont {Cheng}, \citenamefont {Zhang},\ and\ \citenamefont
  {Li}}]{Ke2007}%
  \BibitemOpen
  \bibfield  {author} {\bibinfo {author} {\bibfnamefont {S.~S.}\ \bibnamefont
  {Ke}}, \bibinfo {author} {\bibfnamefont {G.~P.}\ \bibnamefont {Cheng}},
  \bibinfo {author} {\bibfnamefont {L.~H.}\ \bibnamefont {Zhang}},\ and\
  \bibinfo {author} {\bibfnamefont {G.~X.}\ \bibnamefont {Li}},\ }\href
  {https://doi.org/10.1088/0953-4075/40/14/004} {\bibfield  {journal} {\bibinfo
   {journal} {J. Phys. B}\ }\textbf {\bibinfo {volume} {40}},\ \bibinfo {pages}
  {2827} (\bibinfo {year} {2007})}\BibitemShut {NoStop}%
\bibitem [{\citenamefont {Yan}\ \emph {et~al.}(2011)\citenamefont {Yan},
  \citenamefont {Jia}, \citenamefont {Xie},\ and\ \citenamefont
  {Peng}}]{Yan2011}%
  \BibitemOpen
  \bibfield  {author} {\bibinfo {author} {\bibfnamefont {Z.}~\bibnamefont
  {Yan}}, \bibinfo {author} {\bibfnamefont {X.}~\bibnamefont {Jia}}, \bibinfo
  {author} {\bibfnamefont {C.}~\bibnamefont {Xie}},\ and\ \bibinfo {author}
  {\bibfnamefont {K.}~\bibnamefont {Peng}},\ }\href
  {https://doi.org/10.1103/PhysRevA.84.062304} {\bibfield  {journal} {\bibinfo
  {journal} {Phys. Rev. A}\ }\textbf {\bibinfo {volume} {84}},\ \bibinfo
  {pages} {062304} (\bibinfo {year} {2011})},\ \Eprint
  {https://arxiv.org/abs/1201.1163v1} {arXiv:1201.1163v1} \BibitemShut
  {NoStop}%
\bibitem [{\citenamefont {Zhou}\ \emph {et~al.}(2015)\citenamefont {Zhou},
  \citenamefont {Jia}, \citenamefont {Li}, \citenamefont {Yu}, \citenamefont
  {Xie},\ and\ \citenamefont {Peng}}]{Zhou2015}%
  \BibitemOpen
  \bibfield  {author} {\bibinfo {author} {\bibfnamefont {Y.}~\bibnamefont
  {Zhou}}, \bibinfo {author} {\bibfnamefont {X.}~\bibnamefont {Jia}}, \bibinfo
  {author} {\bibfnamefont {F.}~\bibnamefont {Li}}, \bibinfo {author}
  {\bibfnamefont {J.}~\bibnamefont {Yu}}, \bibinfo {author} {\bibfnamefont
  {C.}~\bibnamefont {Xie}},\ and\ \bibinfo {author} {\bibfnamefont
  {K.}~\bibnamefont {Peng}},\ }\href {https://doi.org/10.1038/srep11132}
  {\bibfield  {journal} {\bibinfo  {journal} {Sci. Rep.}\ }\textbf {\bibinfo
  {volume} {5}},\ \bibinfo {pages} {11132} (\bibinfo {year}
  {2015})}\BibitemShut {NoStop}%
\bibitem [{\citenamefont {Pfister}\ \emph {et~al.}(2004)\citenamefont
  {Pfister}, \citenamefont {Feng}, \citenamefont {Jennings}, \citenamefont
  {Pooser},\ and\ \citenamefont {Xie}}]{Pfister2004}%
  \BibitemOpen
  \bibfield  {author} {\bibinfo {author} {\bibfnamefont {O.}~\bibnamefont
  {Pfister}}, \bibinfo {author} {\bibfnamefont {S.}~\bibnamefont {Feng}},
  \bibinfo {author} {\bibfnamefont {G.}~\bibnamefont {Jennings}}, \bibinfo
  {author} {\bibfnamefont {R.}~\bibnamefont {Pooser}},\ and\ \bibinfo {author}
  {\bibfnamefont {D.}~\bibnamefont {Xie}},\ }\href
  {https://doi.org/10.1103/PhysRevA.70.020302} {\bibfield  {journal} {\bibinfo
  {journal} {Phys. Rev. A - At. Mol. Opt. Phys.}\ }\textbf {\bibinfo {volume}
  {70}},\ \bibinfo {pages} {2} (\bibinfo {year} {2004})},\ \Eprint
  {https://arxiv.org/abs/0404049} {arXiv:0404049 [quant-ph]} \BibitemShut
  {NoStop}%
\bibitem [{\citenamefont {Caves}\ and\ \citenamefont
  {Schumaker}(1985)}]{Caves1985}%
  \BibitemOpen
  \bibfield  {author} {\bibinfo {author} {\bibfnamefont {C.~M.}\ \bibnamefont
  {Caves}}\ and\ \bibinfo {author} {\bibfnamefont {B.~L.}\ \bibnamefont
  {Schumaker}},\ }\href {https://doi.org/10.1103/PhysRevA.31.3068} {\bibfield
  {journal} {\bibinfo  {journal} {Phys. Rev. A}\ }\textbf {\bibinfo {volume}
  {31}},\ \bibinfo {pages} {3068} (\bibinfo {year} {1985})}\BibitemShut
  {NoStop}%
\bibitem [{\citenamefont {Schumaker}\ and\ \citenamefont
  {Caves}(1985)}]{Schumaker1985}%
  \BibitemOpen
  \bibfield  {author} {\bibinfo {author} {\bibfnamefont {B.~L.}\ \bibnamefont
  {Schumaker}}\ and\ \bibinfo {author} {\bibfnamefont {C.~M.}\ \bibnamefont
  {Caves}},\ }\href {https://doi.org/10.1103/PhysRevA.31.3093} {\bibfield
  {journal} {\bibinfo  {journal} {Phys. Rev. A}\ }\textbf {\bibinfo {volume}
  {31}},\ \bibinfo {pages} {3093} (\bibinfo {year} {1985})}\BibitemShut
  {NoStop}%
\bibitem [{\citenamefont {You}\ and\ \citenamefont {Nori}(2011)}]{You2011}%
  \BibitemOpen
  \bibfield  {author} {\bibinfo {author} {\bibfnamefont {J.~Q.}\ \bibnamefont
  {You}}\ and\ \bibinfo {author} {\bibfnamefont {F.}~\bibnamefont {Nori}},\
  }\href {https://doi.org/10.1038/nature10122} {\bibfield  {journal} {\bibinfo
  {journal} {Nature}\ }\textbf {\bibinfo {volume} {474}},\ \bibinfo {pages}
  {589} (\bibinfo {year} {2011})},\ \Eprint {https://arxiv.org/abs/1202.1923}
  {arXiv:1202.1923} \BibitemShut {NoStop}%
\bibitem [{\citenamefont {Asb\'oth}\ \emph {et~al.}(2015)\citenamefont
  {Asb\'oth}, \citenamefont {Oroszl\'any},\ and\ \citenamefont
  {P\'alyi}}]{TopInsHun2015}%
  \BibitemOpen
  \bibfield  {author} {\bibinfo {author} {\bibfnamefont {J.~K.}\ \bibnamefont
  {Asb\'oth}}, \bibinfo {author} {\bibfnamefont {L.}~\bibnamefont
  {Oroszl\'any}},\ and\ \bibinfo {author} {\bibfnamefont {A.}~\bibnamefont
  {P\'alyi}},\ }\href {https://doi.org/10.1007/978-3-319-25607-8} {\emph
  {\bibinfo {title} {A Short Course on Topological Insulators: Band-structure
  topology and edge states in one and two dimensions}}},\ Lecture Notes in
  Physics\ (\bibinfo  {publisher} {Springer Cham},\ \bibinfo {year}
  {2015})\BibitemShut {NoStop}%
\bibitem [{\citenamefont {Parto}\ \emph {et~al.}(2020)\citenamefont {Parto},
  \citenamefont {Liu}, \citenamefont {Bahari}, \citenamefont {Khajavikhan},\
  and\ \citenamefont {Christodoulides}}]{Parto2020}%
  \BibitemOpen
  \bibfield  {author} {\bibinfo {author} {\bibfnamefont {M.}~\bibnamefont
  {Parto}}, \bibinfo {author} {\bibfnamefont {Y.~G.}\ \bibnamefont {Liu}},
  \bibinfo {author} {\bibfnamefont {B.}~\bibnamefont {Bahari}}, \bibinfo
  {author} {\bibfnamefont {M.}~\bibnamefont {Khajavikhan}},\ and\ \bibinfo
  {author} {\bibfnamefont {D.~N.}\ \bibnamefont {Christodoulides}},\ }\href
  {https://doi.org/10.1515/nanoph-2020-0434} {\bibfield  {journal} {\bibinfo
  {journal} {Nanophotonics}\ }\textbf {\bibinfo {volume} {10}},\ \bibinfo
  {pages} {403} (\bibinfo {year} {2020})}\BibitemShut {NoStop}%
\bibitem [{\citenamefont {Kato}\ and\ \citenamefont {Aoki}(2015)}]{Kato2015}%
  \BibitemOpen
  \bibfield  {author} {\bibinfo {author} {\bibfnamefont {S.}~\bibnamefont
  {Kato}}\ and\ \bibinfo {author} {\bibfnamefont {T.}~\bibnamefont {Aoki}},\
  }\href {https://doi.org/10.1103/PhysRevLett.115.093603} {\bibfield  {journal}
  {\bibinfo  {journal} {Phys. Rev. Lett.}\ }\textbf {\bibinfo {volume} {115}},\
  \bibinfo {pages} {093603} (\bibinfo {year} {2015})}\BibitemShut {NoStop}%
\bibitem [{\citenamefont {Whalen}(2015)}]{Whalen2015}%
  \BibitemOpen
  \bibfield  {author} {\bibinfo {author} {\bibfnamefont {S.}~\bibnamefont
  {Whalen}},\ }\emph {\bibinfo {title} {{Open Quantum Systems with Time-Delayed
  Interactions}}},\ \href@noop {} {Ph.D. thesis},\ \bibinfo  {school}
  {University of Auckland} (\bibinfo {year} {2015})\BibitemShut {NoStop}%
\bibitem [{\citenamefont {Carmichael}\ \emph {et~al.}(1984)\citenamefont
  {Carmichael}, \citenamefont {Milburn},\ and\ \citenamefont
  {Walls}}]{Carmichael1984}%
  \BibitemOpen
  \bibfield  {author} {\bibinfo {author} {\bibfnamefont {H.~J.}\ \bibnamefont
  {Carmichael}}, \bibinfo {author} {\bibfnamefont {G.~J.}\ \bibnamefont
  {Milburn}},\ and\ \bibinfo {author} {\bibfnamefont {D.~F.}\ \bibnamefont
  {Walls}},\ }\href {https://doi.org/10.1088/0305-4470/17/2/031} {\bibfield
  {journal} {\bibinfo  {journal} {Journal of Physics A: Mathematical and
  General}\ }\textbf {\bibinfo {volume} {17}},\ \bibinfo {pages} {469}
  (\bibinfo {year} {1984})}\BibitemShut {NoStop}%
\bibitem [{\citenamefont {Castellanos-Beltran}\ and\ \citenamefont
  {Lehnert}(2007)}]{Castellanos-Beltran2007}%
  \BibitemOpen
  \bibfield  {author} {\bibinfo {author} {\bibfnamefont {M.~A.}\ \bibnamefont
  {Castellanos-Beltran}}\ and\ \bibinfo {author} {\bibfnamefont {K.~W.}\
  \bibnamefont {Lehnert}},\ }\href {https://doi.org/10.1063/1.2773988}
  {\bibfield  {journal} {\bibinfo  {journal} {Appl. Phys. Lett.}\ }\textbf
  {\bibinfo {volume} {91}},\ \bibinfo {pages} {083509} (\bibinfo {year}
  {2007})},\ \Eprint {https://arxiv.org/abs/0706.2373} {arXiv:0706.2373}
  \BibitemShut {NoStop}%
\bibitem [{\citenamefont {Yeo}\ \emph {et~al.}(2013)\citenamefont {Yeo},
  \citenamefont {Adikan}, \citenamefont {Mokhtar}, \citenamefont {Hitam},\ and\
  \citenamefont {Mahdi}}]{Yeo2013}%
  \BibitemOpen
  \bibfield  {author} {\bibinfo {author} {\bibfnamefont {K.~S.}\ \bibnamefont
  {Yeo}}, \bibinfo {author} {\bibfnamefont {F.~R.~M.}\ \bibnamefont {Adikan}},
  \bibinfo {author} {\bibfnamefont {M.}~\bibnamefont {Mokhtar}}, \bibinfo
  {author} {\bibfnamefont {S.}~\bibnamefont {Hitam}},\ and\ \bibinfo {author}
  {\bibfnamefont {M.~A.}\ \bibnamefont {Mahdi}},\ }\href
  {https://doi.org/10.1007/s00340-012-5260-x} {\bibfield  {journal} {\bibinfo
  {journal} {Applied Physics B}\ }\textbf {\bibinfo {volume} {110}},\ \bibinfo
  {pages} {353} (\bibinfo {year} {2013})}\BibitemShut {NoStop}%
\bibitem [{\citenamefont {Zlobina}\ and\ \citenamefont
  {Kablukov}(2013)}]{Zlobina2013}%
  \BibitemOpen
  \bibfield  {author} {\bibinfo {author} {\bibfnamefont {E.~A.}\ \bibnamefont
  {Zlobina}}\ and\ \bibinfo {author} {\bibfnamefont {S.~I.}\ \bibnamefont
  {Kablukov}},\ }\href {https://doi.org/10.3103/S8756699013040031} {\bibfield
  {journal} {\bibinfo  {journal} {Optoelectronics, Instrumentation and Data
  Processing}\ }\textbf {\bibinfo {volume} {49}},\ \bibinfo {pages} {363}
  (\bibinfo {year} {2013})}\BibitemShut {NoStop}%
\bibitem [{\citenamefont {Berni}\ \emph {et~al.}(2015)\citenamefont {Berni},
  \citenamefont {Gehring}, \citenamefont {Nielsen}, \citenamefont
  {H{\"{a}}ndchen}, \citenamefont {Paris},\ and\ \citenamefont
  {Andersen}}]{Berni2015a}%
  \BibitemOpen
  \bibfield  {author} {\bibinfo {author} {\bibfnamefont {A.~A.}\ \bibnamefont
  {Berni}}, \bibinfo {author} {\bibfnamefont {T.}~\bibnamefont {Gehring}},
  \bibinfo {author} {\bibfnamefont {B.~M.}\ \bibnamefont {Nielsen}}, \bibinfo
  {author} {\bibfnamefont {V.}~\bibnamefont {H{\"{a}}ndchen}}, \bibinfo
  {author} {\bibfnamefont {M.~G.}\ \bibnamefont {Paris}},\ and\ \bibinfo
  {author} {\bibfnamefont {U.~L.}\ \bibnamefont {Andersen}},\ }\href
  {https://doi.org/10.1038/nphoton.2015.139} {\bibfield  {journal} {\bibinfo
  {journal} {Nat. Photonics}\ }\textbf {\bibinfo {volume} {9}},\ \bibinfo
  {pages} {577} (\bibinfo {year} {2015})},\ \Eprint
  {https://arxiv.org/abs/1503.01365} {arXiv:1503.01365} \BibitemShut {NoStop}%
\bibitem [{\citenamefont {Clark}\ \emph {et~al.}(2016)\citenamefont {Clark},
  \citenamefont {Stokes},\ and\ \citenamefont {Beige}}]{Clark2016b}%
  \BibitemOpen
  \bibfield  {author} {\bibinfo {author} {\bibfnamefont {L.~A.}\ \bibnamefont
  {Clark}}, \bibinfo {author} {\bibfnamefont {A.}~\bibnamefont {Stokes}},\ and\
  \bibinfo {author} {\bibfnamefont {A.}~\bibnamefont {Beige}},\ }\href
  {https://doi.org/10.1103/PhysRevA.94.023840} {\bibfield  {journal} {\bibinfo
  {journal} {Phys. Rev. A}\ }\textbf {\bibinfo {volume} {94}},\ \bibinfo
  {pages} {1} (\bibinfo {year} {2016})},\ \Eprint
  {https://arxiv.org/abs/1512.01497} {arXiv:1512.01497} \BibitemShut {NoStop}%
\bibitem [{\citenamefont {Petruccione}\ \emph {et~al.}(2007)\citenamefont
  {Petruccione}, \citenamefont {Breuer},\ and\ \citenamefont
  {Petruccione}}]{Petruccione2007}%
  \BibitemOpen
  \bibfield  {author} {\bibinfo {author} {\bibfnamefont {F.}~\bibnamefont
  {Petruccione}}, \bibinfo {author} {\bibfnamefont {H.~H.-P.}\ \bibnamefont
  {Breuer}},\ and\ \bibinfo {author} {\bibfnamefont {F.}~\bibnamefont
  {Petruccione}},\ }\href
  {https://doi.org/10.1093/acprof:oso/9780199213900.001.0001} {\emph {\bibinfo
  {title} {{The Theory of Open Quantum Systems}}}}\ (\bibinfo  {publisher}
  {Oxford University Press},\ \bibinfo {year} {2007})\BibitemShut {NoStop}%
\bibitem [{\citenamefont {Whalen}\ and\ \citenamefont
  {Carmichael}(2016)}]{Whalen2016}%
  \BibitemOpen
  \bibfield  {author} {\bibinfo {author} {\bibfnamefont {S.~J.}\ \bibnamefont
  {Whalen}}\ and\ \bibinfo {author} {\bibfnamefont {H.~J.}\ \bibnamefont
  {Carmichael}},\ }\href {https://doi.org/10.1103/PhysRevA.93.063820}
  {\bibfield  {journal} {\bibinfo  {journal} {Phys. Rev. A}\ }\textbf {\bibinfo
  {volume} {93}},\ \bibinfo {pages} {063820} (\bibinfo {year} {2016})},\
  \Eprint {https://arxiv.org/abs/1602.03971} {arXiv:1602.03971} \BibitemShut
  {NoStop}%
\end{thebibliography}%

\end{document}